\documentclass[12pt,a4paper,final]{iopart}
\usepackage{iopams}  
\usepackage{graphicx}
\usepackage{amssymb}
\usepackage{amsmath}
\usepackage[breaklinks=true,colorlinks=true,linkcolor=blue,urlcolor=blue,citecolor=blue]{hyperref}

\newcommand {\Be}{\begin{eqnarray*}}
\newcommand {\Ee} {\end{eqnarray*}}
\newcommand {\bey} {\begin{eqnarray}}
\newcommand {\eey} {\end{eqnarray}}
\newcommand{\bit}{\begin{itemize}}      
\newcommand{\eit}{\end{itemize}}
\newcommand{\bfl}{\begin{flusleft}}
\newcommand{\efl}{\end{flusleft}}
\newcommand{\bfr}{\begin{flushright}}
\newcommand{\bc}{\begin{center}}
\newcommand{\ec}{\end{center}}
\newcommand{\ben}{\begin{enumerate}}    
\newcommand{\een}{\end{enumerate}}

\newcommand{\be}{\begin{equation}}
\newcommand{\ee}{\end{equation}}

\renewcommand{\theparagraph}{\Alph{paragraph}}

\def\G{\Gamma}

\def\<{\langle}
\def\>{\rangle}

\newcommand{\crea}[1]{\hat{#1}^{\dagger}}
\newcommand{\op}[1]{\hat{#1}^{}}

\begin{document}

\title[]{Transport of quantum excitations coupled to spatially extended nonlinear many-body systems}

\author{Stefano Iubini$^{1}$}
\ead{stefano.iubini@cnrs-orleans.fr}

\author{Octavi Boada$^2$}
\ead{oboada@lx.it.pt}

\author{Yasser Omar$^{2,3}$}
\ead{yasser.omar@lx.it.pt}

\author{Francesco Piazza$^{1,4}$}
\ead{Francesco.Piazza@cnrs-orleans.fr}

\vspace{7 mm}
\address{$^1$  Centre de Biophysique Mol\'eculaire, (CBM), CNRS-UPR 4301, Rue C. Sadron, 45071, Orl\'eans, France}
\address{$^2$  Physics of Information Group, Instituto de Telecomunica\c{c}\~oes, Lisbon, Portugal}
\address{$^3$  CEMAPRE, ISEG, Universidade de Lisboa, Portugal}
\address{$^4$  Universit\'e d'Orl\'eans, 
               Ch\^ateau de la Source, 45071 Orl\'eans Cedex, France}

\begin{abstract}
The role of noise in the transport properties of quantum excitations is 
a topic of great importance in many fields, from organic semiconductors for 
technological applications to light-harvesting complexes in photosynthesis. 
In this paper we study a semi-classical model where a tight-binding Hamiltonian 
is fully coupled to an underlying spatially extended nonlinear chain of atoms. 
We show that the transport properties of a quantum excitation are subtly 
modulated by (i) the specific type (local vs non-local) of exciton-phonon coupling
and by (ii) nonlinear effects of the underlying lattice. 
We report a non-monotonic dependence of the exciton diffusion coefficient 
on temperature, in agreement with earlier predictions, 
as a direct consequence of the lattice-induced fluctuations in 
the hopping rates due to long-wavelength vibrational modes. A standard measure 
of transport efficiency confirms that both nonlinearity in the underlying lattice 
and off-diagonal exciton-phonon coupling promote transport efficiency at 
high temperatures, preventing the Zeno-like quench observed in other models 
lacking an explicit noise-providing dynamical system.   
\end{abstract}

%

\tableofcontents

\section{Introduction}

Transport of quantum excitations in complex low-dimensional systems is a topic of paramount importance 
in many physical contexts, such as semiconductor nanowires and nanotubes~\cite{Law2004},
metallic wires~\cite{Yanson:1998ly}, ultracold atom systems in one-dimensional 
optical lattices~\cite{Roati2008}, quasi one-dimensional organic superconductors, 
such as semiconducting~\cite{Perebeinos2004} and metal carbon nanotubes~\cite{Wang:2007bh}, 
$\pi$-conjugated polymers~\cite{Zhao:2006qf} and more 
complex quasi-1$D$ nano-architectures for modern technological applications~\cite{Lee:2007zr}, including 
plastic light-emitting devices and organic solar cells~\cite{Scholes2006}.\\
\indent Due to their crucial role in directing light energy to reaction centers during 
the early stages of photosynthesis~\cite{Blankenship:2002ve}, excitons occupy a prominent role among 
the studied quantum excitations~\cite{Amerongen:2000ly}. 
The exciton concept was introduced in solid-state physics by Y. Frenkel in 1931~\cite{Frenkel:1931zr}.
However, it was not until the 1948 seminal paper by A. S. Davydov~\cite{Davydov:1948ys} that this idea was  
applied to geometry-determined molecular systems, such as molecular crystals. 
These and related studies have paved the way for the investigation of exciton transport in light-harvesting 
biomolecules, which contain embedded networks of light-absorbing pigments~\cite{Amerongen:2000ly}.\\
\indent Considerable boost to the investigation of exciton transport in biomolecules has been brought about 
by recent advances in 2$D$ photon echo experiments, which have revealed unusually long decoherence times for excitons 
in light-harvesting complexes~\cite{Engel2007,Read2009,Collini2010,Romero:2014ve} and 
conjugated polymer systems~\cite{Collini2009}. Moreover, theoretical evidence has been accumulated 
that noise in certain regimes could act as a {\em protective} factor for quantum 
coherences~\cite{Chin2013,Chin2010,rebentrost2009environment}, increasing suitably defined 
measures of quantum efficiency related to the transfer of electronic excitation energy from 
a chromophore to a distant one. In turn, such findings have corroborated more detailed investigations 
of the coupled dynamics of exciton transfer and protein vibrations, pointing 
at a functional role of specific vibrational modes in promoting 
possibly function-related, long-lived electronic 
coherences~\cite{OReilly2014,Tiwari2013,Mennucci2011,Plenio:2008fk,Skourtis:2007kx,Lee2007,Wang:2007vn,Adolphs2006}.\\
%
%
%
\indent However, despite the great experimental and theoretical advances, many fundamental questions remain 
open. In particular, pinpointing the {\em structural} determinants of the electron-phonon coupling that 
seem to provide noise-induced protection of exciton dephasing remains  a challenging task. 
Moreover, few studies have addressed the role of the {\em dynamical} determinants of such mechanisms, 
{\em i.e.} the influence of specific inter-atomic and inter-molecular potential energy terms beyond 
the harmonic approximation. Importantly, nonlinearity is known to play a subtle role in many 
quantum transport processes, from heat conduction~\cite{Saito:2007qf} and vibrational 
energy transfer~\cite{Chetverikov:2009ys,Komarnicki:2003uq,Leitner:2001ys} to photon-assisted 
electronic transport in  different nanostructures~\cite{Platero2004}. More generally, 
it is well known that nonlinear effects modulate non-trivially transport in disordered
dynamical systems~\cite{Mulansky:2013zr,Ivanchenko:2011ly,Flach2010,Garcia-Mata2009,Kopidakis2008}.\\
\indent Another issue of paramount importance that needs further investigation 
concerns the details of how the environment ({\em e.g.} the degrees of freedom 
of the protein in photosynthetic complexes) couples to the quantum degrees of freedom. 
In a tight-binding perspective, where the quantum excitation is characterized by a given set 
of site energies $\{\epsilon_{i}\}$ and hopping rates $\{\mathcal{J}_{i}\}$, 
this means investigating the effects of the specific functional dependence 
of $\{\epsilon_{i},\mathcal{J}_{i}\}$  on the degrees of freedom 
of the environment ({\em dynamical} disorder).
For example, letting only site energies fluctuate 
with the environment amounts to adding a pure dephasing term to the unitary evolution 
in the Liouville equation for the time evolution of the one-particle density matrix of the 
quantum excitation. It has been proved that this leads to 
noise-enhanced transfer efficiency~\cite{Chin2010,rebentrost2009environment}. However, 
it has been observed that pure dephasing is not a physically realistic scheme of 
coupling and that in general fluctuations of the hopping rates, even if smaller than 
those of the site energies, can have a considerable impact on the 
dynamics of quantum excitations~\cite{Vlaming:2012qy,Chen:2010uq}. For example, this is the case of
high-mobility organic crystals, such as pentacene and rubrene, where large fluctuations 
in the hopping rates occur at room temperature~\cite{Arago:2015bh,Troisi2011}. 
In this case, it has been shown that Zeno effect at high dephasing rates 
is suppressed and one recovers asymptotic mobility of the 
quantum excitation at increasing noise, albeit
with a diffusion coefficient that decreases with 
temperature~\cite{ScwharzerHaken1972,haken1973exactly,Troisi:2006fk}.\\
\indent Motivated by the above described open questions, in this paper
we adopt a semi-classical modeling strategy to investigate the 
effect of noise on the mobility and transfer efficiency
of a quantum excitation coupled to the vibrations of 
a one-dimensional atomic chain. More  specifically, the 
two main points that we wish to address are: (i) the role of nonlinearity of
the interatomic potentials of the underlying lattice and (ii) 
the role of local versus non-local dynamical disorder. \\
\indent The structure of this paper is as follows. In section~\ref{model} 
we briefly discuss existing modeling strategies and present 
the microscopic model  that we study in this paper, describing a quantum-mechanical 
quasiparticle ({\em e.g.} an exciton or an electron) hopping on a one-dimensional 
lattice~\footnote{Without loss of generality, we will refer to the quantum 
quasiparticle in the following as the {\em exciton}.}.
In section~\ref{results} we present the main
results regarding the spreading properties of an initially localized state in a
thermalized one-dimensional chain.
In section~\ref{sec:efficiency} we analyze the quantum transport
efficiency in our model in the presence of finite local recombination rates.
Finally, in section~\ref{sec:concl} we summarize our conclusions and 
further discuss our main results.

%
%
\section{Exciton-lattice coupled dynamics}\label{model}
%
%
%
%

The time evolution of non-isolated (open) quantum systems, {\em i.e.} systems in contact with a thermal bath,
is an extremely difficult problem in general~\cite{Breuer:2002ve,Rivas:2011qf}.
Several approximate schemes have been proposed, including master equation 
approaches~\cite{schwarzer1972moments,haken1973exactly,Lindblad:1976tg,V.-Gorini:1976hc}
usually based on the projection operator technique~\cite{Nakajima:1958bh},
different non-perturbative methods~\cite{Ishizaki:2009dq,Chen2011,Prior:2010uq} and
path-integral based methods~\cite{Boninsegna:2012kx,Huo:2010uq,Makri:1995cr}.
In some cases the exact solution of the  master equation can be determined analytically~\cite{Kurnosov:2015vn}.\\
\indent Other molecular modeling approaches allow to consider a greater amount of microscopic detail
through ab-initio simulations of both the quantum and the bath ({\em e.g.} the protein) degrees of freedom. 
These techniques combine molecular dynamics simulations for the dynamics of the environment with the time
integration of the Schr\"odinger equation for the reduced quantum system, 
based on quantum electronic structure calculations~\cite{Shim2012,Gutierrez:2009nx,Cheung:2009oq,Viani:2014kl,Mennucci2011}.
Such methods, often referred to as Quantum Mechanics/Molecular Mechanics (QM/MM) are more 
sophisticated and detailed modeling schemes belonging to a more general family of modeling strategies,
where the full quantum evolution of the system is parametrized by the classical coordinates 
of the underlying lattice/protein, that evolves in parallel according to Newton 
equations. Such schemes have been applied with success to a variety of problems, 
including energy and charge transport in polypeptide chains~\cite{Hennig2001} 
and other biomolecules~\cite{Komarnicki:2003uq,Mingaleev1999,Bittner2010,Freedman:2010kx}.\\
\indent In the tight-binding approximation, the most general Hamiltonian governing the
propagation of an exciton coupled to a one-dimensional lattice can be written as
\begin{equation}
\label{firstham}
H =   \sum_{nm} J_{nm} \left( 
                              u_n, u_m 
                       \right)  B^{\dagger}_n B_m
    + \sum_{m} \frac{p^2_n}{2M} + 
      \sum_{mn} V \left( 
                    u_n,u_m 
                  \right) , 
\end{equation}
where $M$ is the bead mass, $B^{\dagger}_n$ is an exciton creation operator at
site $n$, and $u_n$ is the displacement of the $n$-th mass with respect to 
its equilibrium position. The term $V\left(u_n,u_m\right)$ represents the total potential energy of the 
nodes $m$ and $n$ including possible onsite terms.  The energies $J_{nm} \left( u_n, u_m  \right)$
describe the modulation of the quantum Hamiltonian due to the fluctuations of the
underlying chain. In principle, depending on the physical nature of the system, 
both the site energies ($n=m$ terms) and 
the hopping rates ($n\neq m$ terms) can be be influenced by the vibrations of the chain atoms.  \\
\indent The Hamiltonian in Eq.~\eref{firstham} describes a general class of semi-classical 
models, including the Davydov~\cite{Davydov:1962vn,Scott1992}, the 
Holstein~\cite{Holstein1959,Kalosakas:1998bs}, and the Su-Schrieffer-Heeger (SSH)~\cite{Su:1979ij} 
Hamiltonians, that have been employed to describe the dynamics of 
different kinds of quantum excitations in a variety of physical systems.
In such modeling schemes, one treats the exciton as a quantum mechanical
particle, while describing the oscillations of the lattice classically. Thus,
the equations of motion (EOM)  are given by a time-dependent effective Hamiltonian for the
exciton, which  depends on the lattice variables $u_n$. The EOMs for the
chain are those of a set of coupled oscillators driven by the exciton wave
function.  \\
\indent In this paper we specialize to a nearest-neighbour tight-binding scheme with 
fully fluctuating parameters ({\em i.e.} both site energies and hopping rates), coupled
to a nonlinear Fermi-Pasta-Ulam (FPU) chain~\cite{E:1955hs}.
The FPU potential can be derived as a Taylor expansion of a generic nearest-neighbour
interaction potential with respect to the equilibrium positions of the chain.
However, for the sake of simplicity we neglect cubic terms, which are known to 
give rise to specific topological (kink) excitations in the system, where also the equilibrium position
of atoms are shifted and also to 
more complex combined breather-kink modes~\cite{Butt:2007aa}.\\
\indent The Hamiltonian governing the system 
is the sum of two terms $H=H_e + H_{l}$. $H_e$ is the Hamiltonian for an exciton propagating
on the lattice in a given dynamical configuration, while
$H_l$ is the lattice Hamiltonian. Explicitly, we have 
\begin{equation}
H_e=   \sum_{n=1}^{L} \epsilon_n(u) B^{\dagger}_n B_n 
     + \sum_{n=1}^{L} J_n(u) \left( 
                          B^{\dagger}_{n+1} B_n + B^{\dagger}_{n} B_{n+1}
                     \right)                                      
\end{equation}
and
\be
\label{eq:H_latt}
H_l = \sum_{n=0}^{L} \left[ 
                          \frac{p_n^2}{2M}  +  \frac{\kappa}{2}(u_{n+1}-u_n)^2 +
                          \frac{\beta}{4} (u_{n+1}-u_n)^4 
                     \right] 
\ee
where $L$ is the lattice size, $M$ is the mass of the atoms and $\beta$ is the
anharmonicity parameter. The operator $B_n$ ($B^{\dagger}_n$) destroys (creates)
an exciton at the position occupied by the $n$-th bead.  \\
\indent According to our scheme, the energies appearing 
in the exciton hamiltonian $H_e$ are renormalized over time by the lattice fluctuations, 
namely
\be\label{renormenergy}
\epsilon_n = E_n^0 + \chi_E (u_{n+1}-u_{n-1})\, ,
\ee
and
\be\label{renormJ}
J_n = J_n^0 + \chi_J (u_{n+1}-u_n) \quad\,.
\ee
In the above equations, the $0$ superscript refers to the unperturbed values of
site energies and hopping integrals, while  the parameters $\chi_E$ and $\chi_J$ 
gauge the strength of  the exciton-lattice coupling. In the following we
will consider $E_n^0=J_n^0=1$ $\forall$ $n$, unless specifically stated otherwise.
Notice that with the above choice of coupling the total Hamiltonian is invariant under global spatial
translations of the $u_n$, so  that the total momentum of the lattice is conserved during the evolution.
Moreover, the invariance of the Hamiltonian under phase transformations of the
variables $\op{B}_n$  guarantees the  conservation
of the total excitonic probability. Eqs.~\eqref{renormenergy} and~\eqref{renormJ} 
are a natural way to minimally couple the exciton to the chain, as they can be
seen as the first order term of a Taylor expansion of the effective local
energies and exchange integrals in powers of $u_n$.\\
\indent In our semi-classical approach, only the exciton state is treated
quantum-mechanically, while the lattice variables $u_n$ evolve according to 
the laws of classical mechanics.
As we work in the single-exciton manifold, in order to determine
the time-evolution of the system we consider a trial
wave-function for the exciton in the form
\be
|\psi(t)\>=\sum_n b_n^*(t)\ \crea{B}_n|0\>
\ee
and derive the appropriate coupled EOMs for the coefficients $b_n$ and lattice variables
$u_n$. The amplitudes $b^*_n$ define the wavefunction in the basis of lattice
sites. In the following we will always impose the normalization condition 
$\sum_n |b_n|^2=1$. The time evolution 
of the coefficients $b_n$ follows directly from Schr\"odinger's equation, 
\be
\label{EOMexciton}
i\hbar\frac{{\rm d} b_n}{{\rm d} t} = - \epsilon_n b_n 
                                      - J_n (b_{n+1} + b_{n-1})
\ee
where the parameters $J_n$ and $\epsilon_n$ are those of Eqs.~\eqref{renormenergy}
and~\eqref{renormJ}, and depend on the lattice variables. The EOMs of the underlying lattice
are obtained from the expectation value of the Hamiltonian on the exciton wavefunction, that is
\be
\label{EOMnetwork}
\begin{aligned}
M \ddot{u}_{n} &= -\frac{\partial}{\partial u_n} \langle \psi | H | \psi \rangle \\
               &= F^{l}_{n} + F^{e}_{n} 
\end{aligned}
\ee
with
\begin{equation}
\begin{aligned}
\label{e:forces}
F^{l}_{n} &= \kappa (u_{n+1} + u_{n-1} - 2u_n)  +  
             \beta \left[ 
                          (u_{n+1} - u_n)^3 + 
                          (u_{n-1} - u_n)^3 
                    \right] \\
F^{e}_{n}  &= 2 \chi_J \textrm{Re}( b^{*}_{n+1}b_{n}  - b^{*}_{n-1}b_{n}) +
              \chi_E (b^{*}_{n+1}b_{n+1} - b^{*}_{n-1}b_{n-1})
\end{aligned}
\end{equation}
In the following we will take $\kappa=M=1$ so that the 
upper frequency of the linear spectrum of the chain is $\omega_{0} = 2$.
Moreover, throughout this paper we consider periodic boundary conditions. 
It is worth noting that Eqs.~\eqref{EOMexciton} and~\eqref{EOMnetwork} also correspond 
to the deterministic version of a set of equations that can be derived 
through a path-integral approach~\cite{Boninsegna:2012kx}. Furthermore, we observe 
that, despite our choice of a bilinear exciton-lattice coupling, the ensuing time evolution 
governed by Eqs.~\eqref{EOMexciton} and~\eqref{EOMnetwork} is a nonlinear one.\\
\indent For the analyses reported in the following, 
we evolved numerically the coupled EOMs~\eqref{EOMexciton} 
and~\eqref{EOMnetwork} starting from different initial conditions and referring both to equilibrium 
and non-equilibrium setups. 
For this we used a standard $4-th$ order Runge-Kutta algorithm with a time step $dt=10^{-3}$.
In particular, we prepared the lattice initial condition by sampling 
a characteristic configuration of the variables $u_n$ and $p_n$ representative  of a finite 
temperature $T$. This task is accomplished by
thermalizing the lattice via a Langevin heat bath~\cite{LLP03} at temperature $T$ for a sufficiently 
long transient time $t_0$.
Explicitly, the Langevin thermalization is achieved by augmenting the free lattice EOMs
with a suitable friction term and a stochastic force, namely
\be
M \ddot{u}_{n} =  F^{l}_{n}  -\gamma p_n + \sqrt{2\gamma T} \xi_n (t)
\ee
where $\gamma$ defines the coupling strength of the reservoir and $\xi_n (t)$ is a Gaussian white noise with zero
mean and unit variance.
Notice that the present setup corresponds to the thermalization of each site of the lattice chain with an
independent heat bath at temperature $T$ (measured setting the Boltzmann constant $k_B=1$).
Moreover, for all the simulations we will set $t_0=L$ and $\gamma=1$.
This choice of parameters guarantees an efficient thermalization of the lattice in the whole range of parameters 
studied in this paper. 
Finally, once the lattice has reached a stationary, thermalized state, the Langevin reservoir is 
disconnected in order to sample the microcanonical dynamics of the total (lattice + exciton) system.





%
\subsection{Equilibrium spectral analysis}
%

\begin{figure}[t!]
\begin{center}
\includegraphics[width=0.9\textwidth,clip]{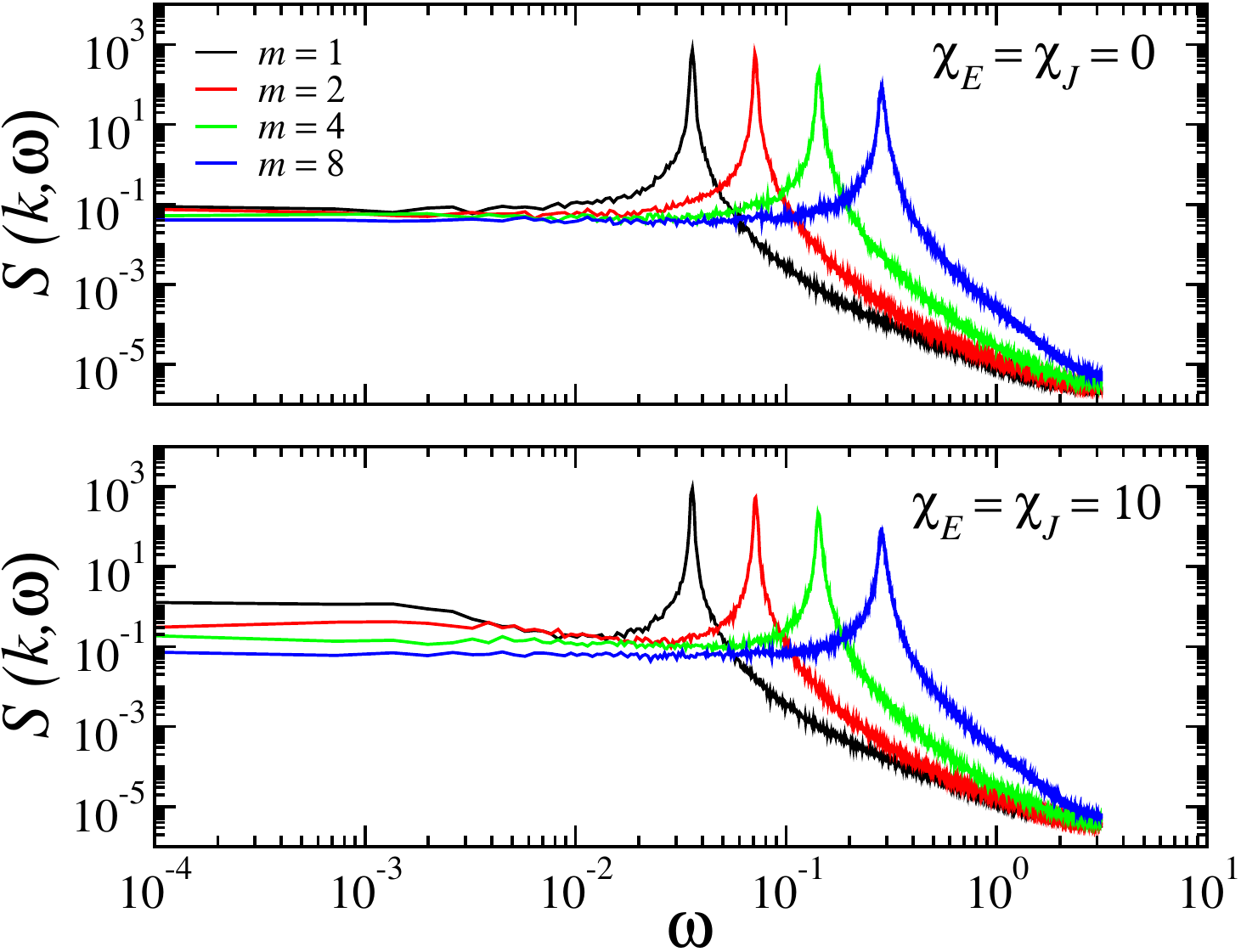}
\caption{(Color online) Equilibrium power spectra  $S(k,\omega)$  of the lattice dynamics 
(power spectrum of the time series $\sum_{n}u_{n}(t) e^{ikn}$) for low-$k$ Fourier 
modes with $k=2\pi m/L$ and $m=\{1,2,4,8\}$ for a thermalized nonlinear chain with $T=1$, 
$\beta=1$ and $L=256$.
The upper panel refers to an isolated lattice, while the lower panel shows the lattice 
spectrum in a regime of a strong coupling with the exciton.
Each spectrum is computed by averaging over $50$ independent realizations of the dynamics.} 
\label{f:fig0a}
 \end{center}
\end{figure}

\begin{figure}[t!]
\begin{center}
\includegraphics[width=0.9\textwidth,clip]{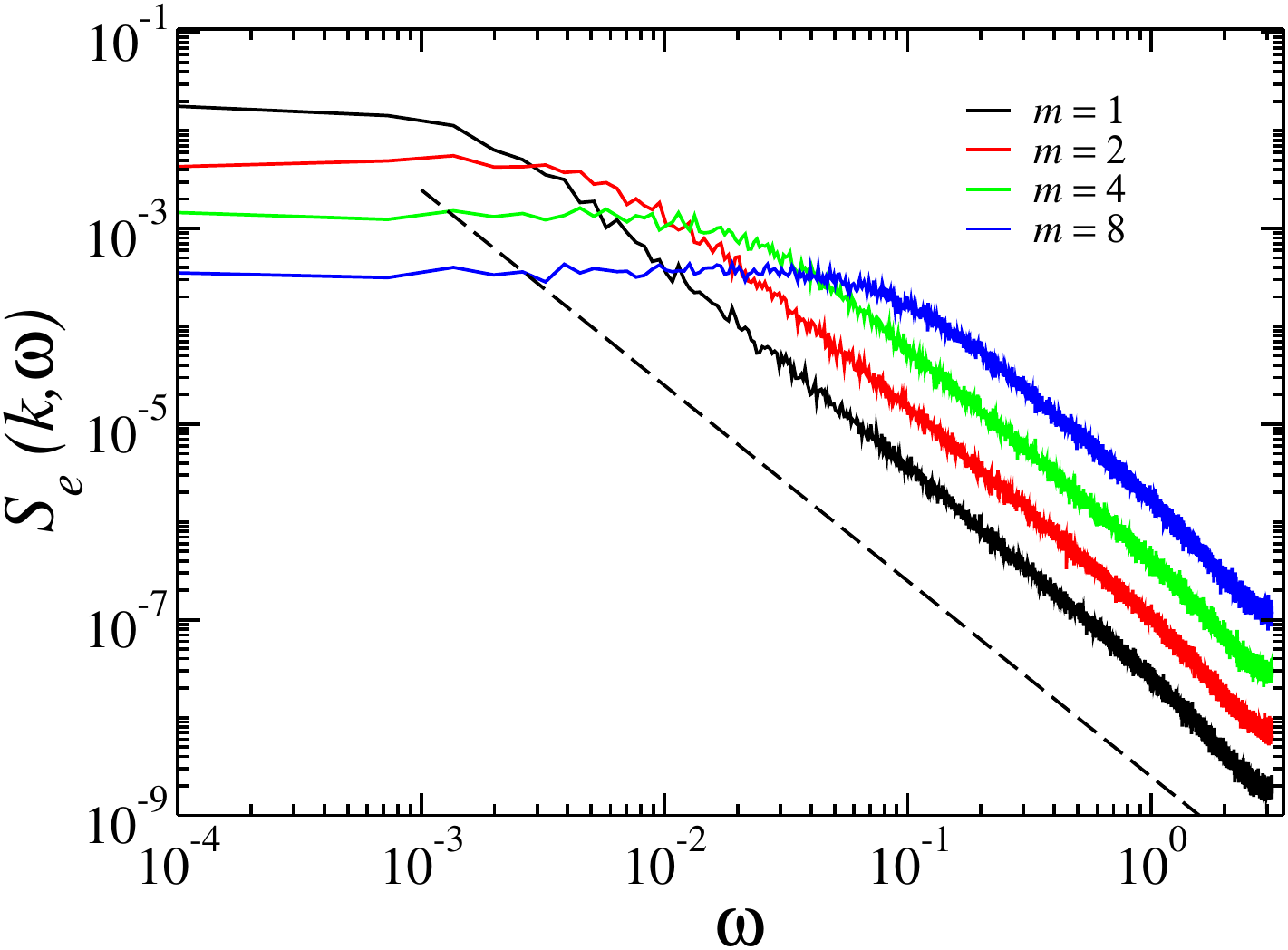}
\caption{(Color online) Power spectra $S_e(k,\omega)$ of the exciton amplitudes 
({\em i.e.} power spectra of the time series $\sum_{n} |b_{n}(t)|^{2} e^{ikn}$)
for an exciton-lattice coupling $\chi_{E} = \chi_{J} = 1$ in a chain with $T=1$, 
$\beta=1$ and $L=256$ (with $k=2\pi m/L$).
Each spectrum is computed by averaging over $50$ independent realizations of the 
dynamics of the underlying lattice. The dashed black line is a power-law with exponent $-2$ and 
indicates the characteristic diffusive behavior.}
\label{f:fig0b}
 \end{center}
\end{figure}

The explicit treatment of the lattice dynamics by means of Eq.~\eqref{EOMnetwork}
represents a simple and direct way to include more realistic, {\em explicit} spatio-temporal correlations 
in the environmental noise that perturbs the exciton evolution, as opposed to more abstract 
treatments where the environment only enters the picture as a {\em spectral density}. 
It is therefore interesting to provide a spectral characterization of the lattice 
dynamics, focusing on equilibrium stationary states at a given temperature $T$.
The same analysis also allows us to explore the near-equilibrium exciton transfer processes.\\
\indent In order to show the relevant transport properties of the lattice, it is instructive to look 
at the power spectra of its long-wavelength Fourier modes. These are reported in the 
upper panel of Fig.~\ref{f:fig0a}.
Simulations have been performed evolving the system in the presence of a Langevin thermal bath 
at temperature $T$~\cite{LLP03} for a transient time $t_0=L$.
We remark that, according to our 
prescriptions, time in our simulations is measured in units of $\sqrt{M/\kappa}$. 
The external heat bath was switched off at $t = t_0$ and the power spectrum was then 
computed by sampling the microcanonical dynamics over an interval of $2^{16}$ 
temporal units and averaging over different  thermalized initial conditions.
\\
\indent For a white-noise signal, one would obviously observe a flat spectrum.
The sharp peaks that are visible in Fig.~\ref{f:fig0a} flag the nontrivial propagation 
of correlations inside the nonlinear chain.
Their presence is also closely related to the anomalous heat transport properties observed in
the FPU chain~\cite{LLP03} and clearly shows that the lattice dynamics cannot be approximated by
a diffusive uncorrelated process like a pure-dephasing noise. We conclude that nonlinearity 
in the interatomic potentials of the underlying dynamical system couples the exciton 
to a noise possessing a complex structure. This is likely to be the case {\em a fortiori} 
for excitons propagating within complex fluctuating biomolecules. The necessity of including 
the proper correlations in the noise beyond pure-dephasing, 
possibly encoded in the underlying lattice structure, appears therefore important.\\
%
%
\indent We then move on to examine the lattice spectrum in the presence of
non-vanishing  exciton-lattice coupling, as this can inform on the back-action exerted 
by the exciton on the underlying dynamical system.  
In this situation, the thermalization process and the following free dynamics were performed 
by evolving the coupled equations~\eqref{EOMexciton} and~\eqref{EOMnetwork} with an excitonic initial
condition corresponding to a random-phase delocalized state.  
The corresponding lattice spectra are shown in the lower panel of Fig.~\ref{f:fig0a}. 
We find that,  even in the regime of strong coupling,  the 
relevant features of the characteristic peaks are essentially unchanged. The only difference
with the zero-coupling case is a slight deformation of the low-frequency and 
low-wavenumber region of the spectrum.
We therefore conclude that our combined exciton-lattice model exhibits environmental correlations that 
generally survive for finite temperatures and coupling strengths.\\
\indent It is instructive to carry out a similar spectral analysis also for the excitonic degrees of freedom. 
In Fig.~\ref{f:fig0b} we show the power spectrum $S_e(k,\omega)$ of the exciton
amplitude field $|b_n(t)|^2$ in the presence of a thermal background at finite temperature.
The spectrum of each normal mode $k$ is well fitted by a Lorentzian distribution, a manifest
evidence  that sufficiently strong  perturbations resulting from the underlying lattice dynamics 
produce a diffusion-dominated exciton transport.
%
%
It is important to observe that the spectral analysis  reported
here  allows to explore the out-of-equilibrium properties of the exciton-lattice chain  
only perturbatively, {\em i.e.} in the spirit of linear-response theory. 
On the other hand,  many realistic situations are characterized by strong out-of-equilibrium conditions.
For example,  this is the case of the propagation of photosynthetic 
excitons created in light-harvesting antenna complexes following the absorption of a photon. 
Such physical scenarios properly correspond to far-from-equilibrium initial condition 
that cannot be included in linear-response schemes.
For this reason, in the next section we discuss in detail  non-stationary 
exciton transport arising from spatially localized initial conditions.

%
\section{Exciton spreading on a chain at finite temperature}\label{results}
%

In this section we study the spreading of an initially localized exciton
wave-function interacting with the chain. The initial conditions for the lattice 
are taken by sampling random velocities from a Maxwell distribution at a given temperature $T$, 
which is one of the parameters in these simulations. The initial displacements are set to zero.
The lattice is then evolved for a transient time $t_0=L$ in the presence of a Langevin heat bath
at temperature $T$ that interacts independently with each site of the chain~\cite{LLP03}. After this 
thermalization process, the external heat bath is disconnected and the  lattice-exciton 
interaction term in the Hamiltonian is switched on.
Furthermore, we average the time evolution of the same initial condition for the exciton 
over many independent trajectories, each corresponding to different initial 
conditions of the chain sampled from the same thermal distribution. 
In our picture, the decoherence of the exciton wave-function is brought about by 
averaging over many independent realizations of the explicit noise (the lattice dynamics).\\
\indent For sufficiently low temperatures and couplings, one can argue that
the exciton evolution should be only weakly perturbed by the environment. Indeed, 
the exciton is found to spread over the chain almost ballistically in this regime,  with a
slow loss of coherence. However, one may speculate that for larger values of couplings and
temperatures  the non-Markovian nature of the noise acting upon the exciton  and
the nonlinearity of the dynamics may play a fundamental role in modulating the spreading 
of an exciton. For instance, very large couplings typically result in the emergence of immobile  
self-trapped states of nonlinear origin. The net effect is that a substantial amount 
of the vibrational energy gets pinned around a handful of sites in the chain, producing 
a pinning potential where the exciton self-traps, 
making {\em de facto} impossible any kind of exciton transport.
Self-trapped states are well-known in many problems studied with models belonging 
to the same class as ours, such as the Holstein polaron~\cite{Kalosakas:1998bs,Holstein1959}
and discrete breathers (DB) in dilute Bose-Einstein condensates trapped in optical 
potentials~\cite{Fleischer:2003fv,Franzosi:2011dz,Flach2008,Trombettoni2001}.\\
\indent The situation is perhaps more interesting at intermediate couplings, where 
it is not clear {\em a priori} over what timescale the spreading is diffusive 
and what is the dependence of the exciton diffusion constant $D$ on the lattice temperature.\\
\indent In order to gather information on the fraction of lattice sites that are
significantly occupied during the time evolution of the system, 
we compute the participation ratio $\Pi$, defined as
\be
\Pi(t) =\frac{\sum_{n} |b_n(t)|^2}{\sum_{n} |b_n(t)|^4}-1= 
\frac{1}{\sum_{n} |b_n(t)|^4}-1\quad.
\ee
With this choice of normalization, it is easy to show that $\Pi=0$ for a completely
localized exciton wavefunction, while one has $\Pi = L-1$ for a perfectly uniform state.
Therefore, one should think at $\Pi$ as an effective {\em length}, measuring 
the spatial extent of the exciton wavefunction over the chain. As such, diffusive 
spreading would be flagged by a law of the type $\Pi(t)\propto t^{1/2}$, while 
ballistic propagation would correspond to a linear dependence on time, 
$\Pi(t)\propto t$.  \\
%
%
\begin{figure}[t!]
\begin{center}
\includegraphics[width=0.7\textwidth,clip]{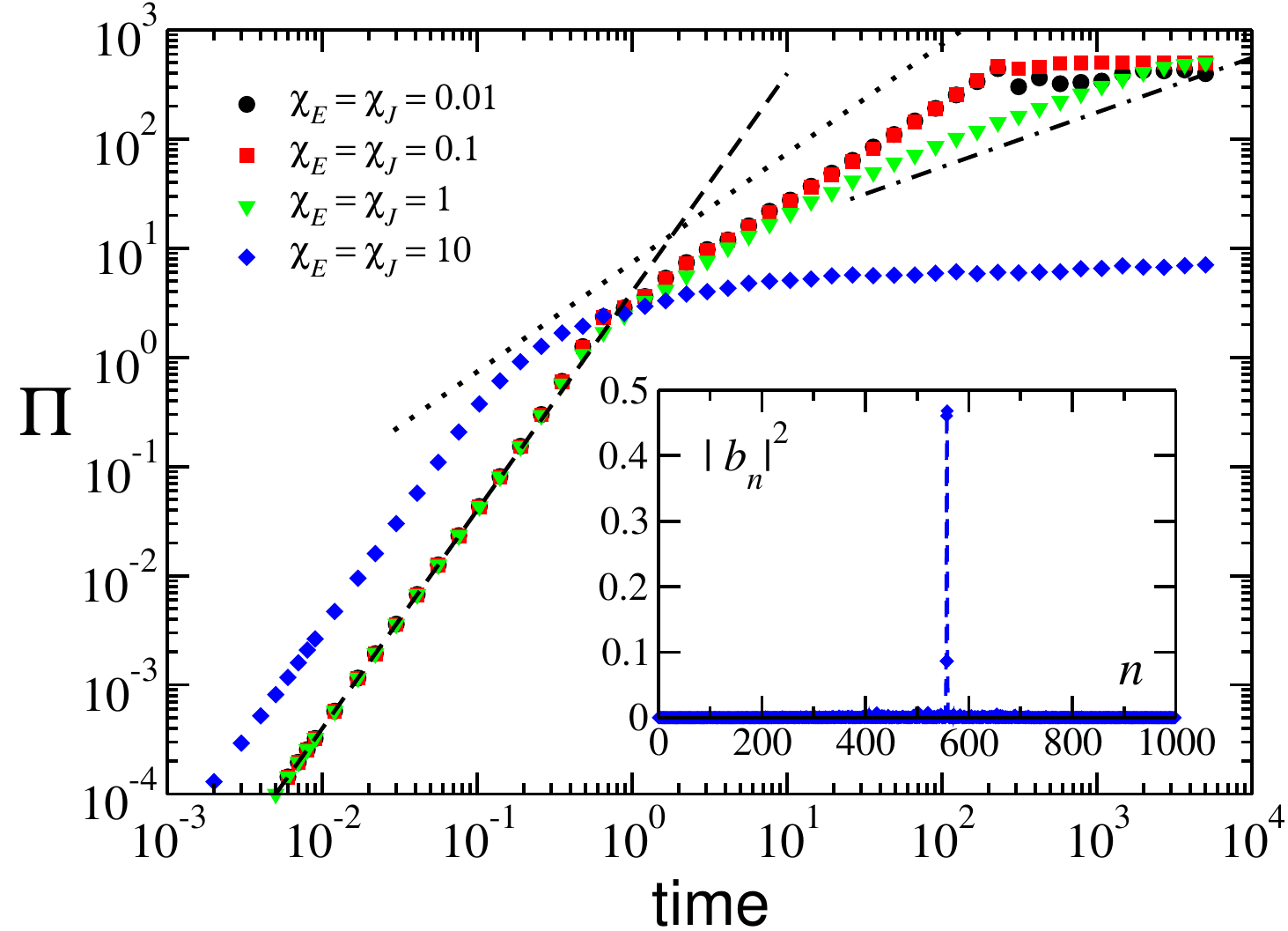}
\caption{(Color online) Evolution of $\Pi(t)$ for $T=0.1$ and different couplings  
in a chain with $L=1000$ lattice sites. The  dashed line is a plot of the 
initial super-ballistic propagation $\Pi(t) = (t/\tau)^{2}$ with $\tau=\hbar/(2J)$.
Dotted and dot-dashed lines indicate a power-law growth of the kind $t^{\alpha}$
with $\alpha=1$ (ballistic) and $0.5$ (diffusive), respectively. 
The inset shows the final profile of $|b_n|^2$ for the largest 
coupling $\chi_{E}= \chi_{J}=10$.  A self-trapped state is clearly observable 
around the center of the chain.}
\label{f:fig1}
\end{center}
\end{figure}
%
\indent In Fig.~\ref{f:fig1} we show the typical time evolution of the participation
ratio for different  coupling strengths and $T=0.1$. 
The first stage is a short transient $(t\lesssim 1)$, where $\Pi$ grows quadratically in time.
This {\em super-ballistic} evolution is characteristic of the very first stage of the time
evolution of an initially localized wavefunction. This can be easily proved by writing 
down Eqs.~\eqref{EOMexciton} for an exciton initially sitting entirely at site $n$
in a chain with $J_{n} = J$ $\forall \ n$. 
In this case it is not difficult to show that, if one defines 
$Q(t) = |b_{n}(t)|^{2} - |b_{n\pm 1}(t)|^{2}$ (with $Q(0)=1)$, then
\be
\label{e:dimer}
Q(t) = \exp \left[
                     - \frac{1}{\tau^{2}}
                       \int_{0}^{t} \sin 2\Delta \phi \, 
                       \Delta \dot{\phi}^{-1} \, dt^{\prime}         
             \right]
\ee
where $\tau=\hbar/(2J)$ and $\Delta \phi = \phi_{n\pm1}-\phi_{n}$ 
is the phase difference between the initially excited site and its neighbors. 
Without loss of generality, in the spirit of a Taylor expansion, 
we can assume $\Delta \phi \simeq t/\tau$ in the early 
stages of the propagation. It is then straightforward to show that this immediately leads to
a super-ballistic trend, namely $\Pi(t) \simeq (t/\tau)^{2}$.
It is apparent from Fig.~\ref{f:fig1} that this prediction is in excellent agreement 
with the simulations for low and moderate couplings. For larger values of the coupling,
the physical law does not appear to change, while the time constant turns out to be renormalized.
For example, for $\chi_{E}=\chi_{J}=10$ we get $\Pi(t) \simeq (t/\tau^{\prime})^{2}$
with $\tau^{\prime} = \tau/3$, which means faster super-ballistic propagation for large coupling strengths.
%
%
It is intriguing to observe that this first super-ballistic stage could be physically relevant 
in many contexts. For example, in light-harvesting complexes one typically has $J \simeq 100$ cm$^{-1}$,
{\em i.e.} $\tau \simeq 170$ fs, which is of the same order of magnitude as the observed 
lifetime of quantum beats in 2$D$ photon echo spectroscopy at room temperature~\cite{Engel2007}.\\
\indent In the subsequent evolution for times greater than $\tau=\hbar/(2J)$,
we can single out two main regimes corresponding to different dynamical situations. 
At low couplings, we recover the expected almost unperturbed evolution associated 
with ballistic spreading of the exciton, {\em i.e.} $\Pi(t) \propto t$. 
The plateau observed at long times in Fig.~\ref{f:fig1} for $\chi_{E}=\chi_{J}=0.01$ and 0.1
simply signals that the exciton wavefunction has reached complete delocalization 
(the lattice is finite) and no further spreading is thus possible. 
For intermediate couplings, it is clearly possible to observe a transition from a ballistic to 
a diffusive ($\Pi(t) \propto t^{1/2}$) regime.
Overall, we conclude that for intermediate couplings one should always expect a ballistic-to-diffusive 
crossover, the time scale associated with it decreasing with increasing 
coupling.\\
\indent The situation changes rather dramatically  for large values of the coupling strengths. 
In this case, the system enters a strongly nonlinear 
regime, where the localized initial condition triggers 
the spontaneous creation of a stable self-trapped state of nonlinear origin,
which results in a very slow sub-diffusive transport
(see also the inset of Fig.~\ref{f:fig1}). Increasing the coupling further causes the initial
amplitude of the exciton to stay permanently stored in a localized, time-periodic excitation of the system
which is virtually decoupled from the rest of the system
(cf. Ref.~\cite{Flach2008} for a comprehensive review on discrete breathers in nonlinear lattices).
It is important to point out that the asymptotic stability of the
self-trapped state depends both on the strength of the coupling $\chi_{E}=\chi_{J}$ and 
on the temperature $T$. In particular,  for certain critical values $\chi_c$ and $T_c$, 
such localized structures become unstable and get quickly destroyed by the thermal fluctuations of the 
lattice~\cite{Davydov:1979,Cruzeiro-Hansson:1995hc}. On the other hand, the 
ballistic and diffusive regimes shown in Fig.~\ref{f:fig1} are not separated by 
a true dynamical transition. In fact, one can argue that for sufficiently long 
chains and times, any arbitrarily small interaction with the lattice will 
eventually cause a diffusion of the excitonic wavepacket.\\
\indent For the exciton nonequilibrium evolution reported in Fig.~\ref{f:fig1} 
we have also monitored the lattice kinetic temperature $T_k$ {\it measured} 
after the Langevin heat bath has been disconnected. $T_k$ is defined as 
\be
\label{eq:kin_temp}
T_k(t)=\frac{1}{L}\sum_{n=1}^L \left[\langle p_n^2(t) \rangle - \langle p_n(t) \rangle^2\right]
\ee
where the symbol $\langle \cdot \rangle$ refers to a classical average over the set of 
independent lattice  trajectories.
%
\begin{figure}[ht]
\begin{center}
\includegraphics[width=0.7\textwidth,clip]{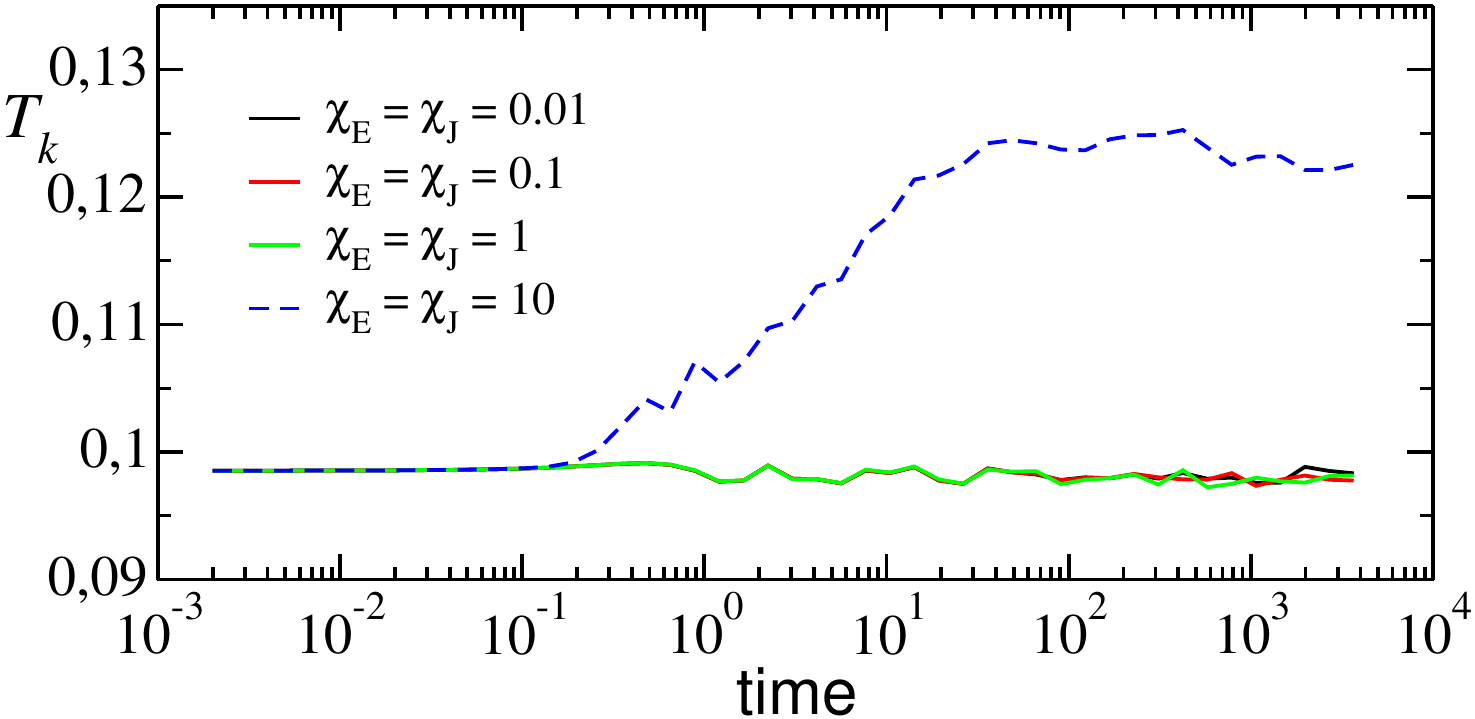}
\caption{Evolution of the lattice temperature $T_k$ (see Eq.~\ref{eq:kin_temp}) 
during the exciton spreading dynamics shown in Fig.~\ref{f:fig1}.}
\label{f:latt_temp}
\end{center}
\end{figure}
%
Fig.~\ref{f:latt_temp} illustrates the evolution of $T_k$ for the same setup 
of Fig.~\ref{f:fig1}, which corresponds to a  temperature of the Langevin bath 
$T=0.1$. Interestingly, in a wide region of coupling values that keep 
the overall system out of the strongly nonlinear regime,  
$T_k$  remains close to the Langevin temperature during  the whole exciton evolution. 
Conversely, the emergence of a stable discrete  breather for large coupling strengths,
$\chi_E=\chi_J=10$ (cf. blue diamonds  of Fig.~\ref{f:fig1}),
produces a clear increase of $T_k$, which is associated with the
conversion of a substantial portion of lattice energy into (negative) 
exciton-lattice interaction energy. Accordingly, by virtue of the conservation of the total
energy of the system, such energy transfer causes the lattice to heat up. 
However, this should be regarded as a finite-size effect. 
In general, we expect that,  upon increasing the lattice 
length $L$, the heating effect becomes less and less  important  until it should eventually 
disappear in the thermodynamic limit $L\to\infty$,
since the breather interaction energy is localized over a finite number 
of lattice sites, whereas the lattice energy scales linearly with $L$.
Altogether, the above analysis confirms the consistency of the lattice 
dynamics as a well defined explicit thermal environment.\\
%
\subsection{The effective diffusion constant}

\begin{figure}[ht]
\begin{center}
\includegraphics[width=0.7\textwidth,clip]{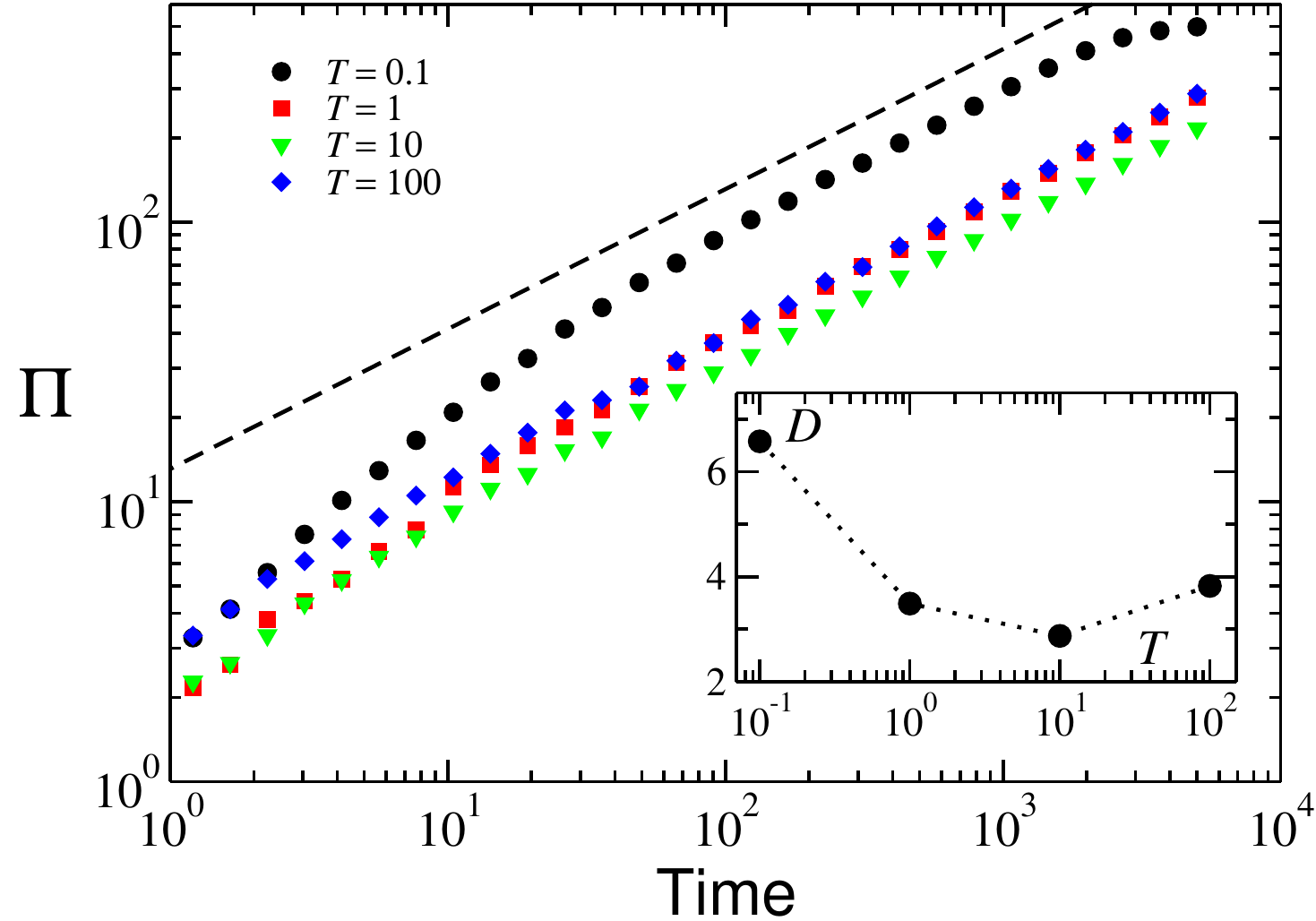}
\caption{(Color online) Time evolution of $\Pi(t)$ for $\chi_{E}=\chi_{J}=1$ 
and different temperatures of the lattice. The inset shows the behavior of the diffusion 
constant measured by fitting a power law to the asymptotic portion of the spreading.
The dashed line is a plot of a power law with exponent 1/2.}
\label{f:fig2}
 \end{center}
\end{figure}

\begin{figure}[ht]
\begin{center}
\includegraphics[width=0.7\textwidth,clip]{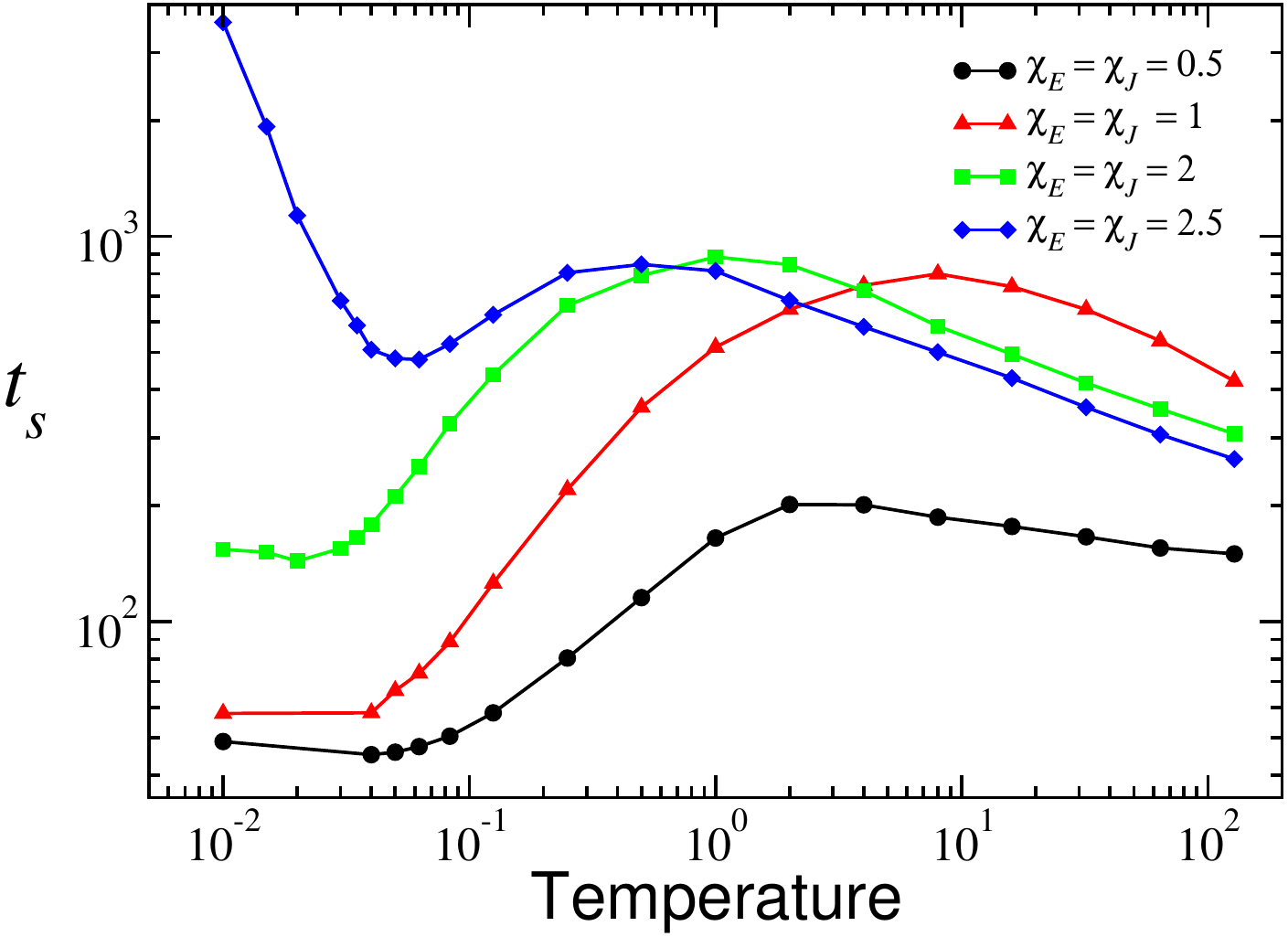}
\caption{(color online) The average time $t_s$ needed to attain a participation 
ratio $\Pi_{s} = 100$ as a function of temperature $T$ for different  values of the
coupling strengths. The chain length is $L=1000$.}
\label{f:fig3}
 \end{center}
\end{figure}

\begin{figure}[t!]
\begin{center}
\includegraphics[width=0.7\textwidth,clip]{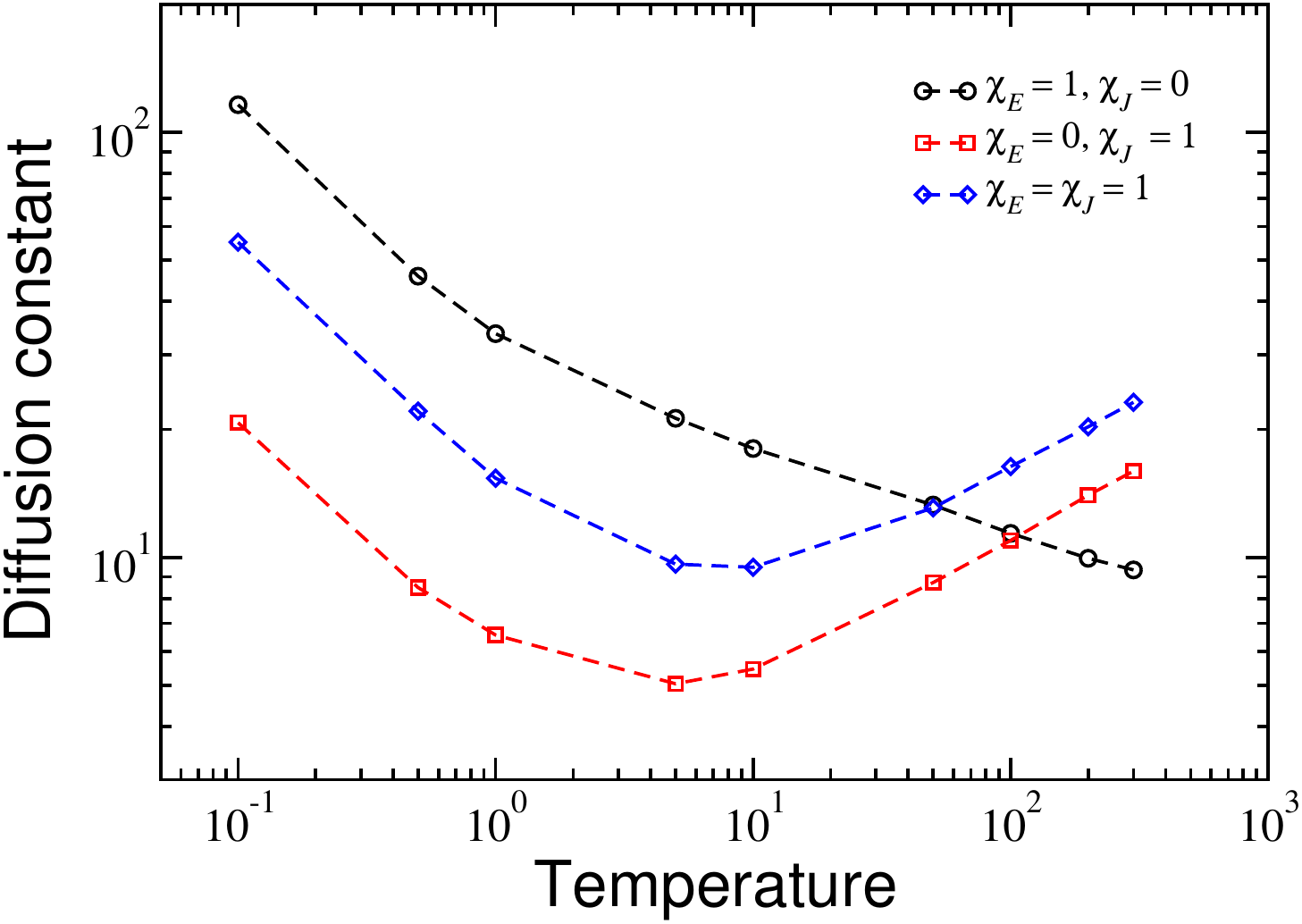}
\caption{(Color online) Temperature dependence of the spreading diffusion constants $D$ in the 
presence of purely diagonal interaction (black circles) and purely off-diagonal 
interaction (red squares). Blue diamonds refer to the {\em full} 
model, with $\chi_E=1$ and  $\chi_J=1$.  Other parameters are $L=1000$ and $\beta=1$.}
\label{f:fig4}
\end{center}
\end{figure}

\begin{figure}[t!]
\begin{center}
\includegraphics[width=0.7\textwidth,clip]{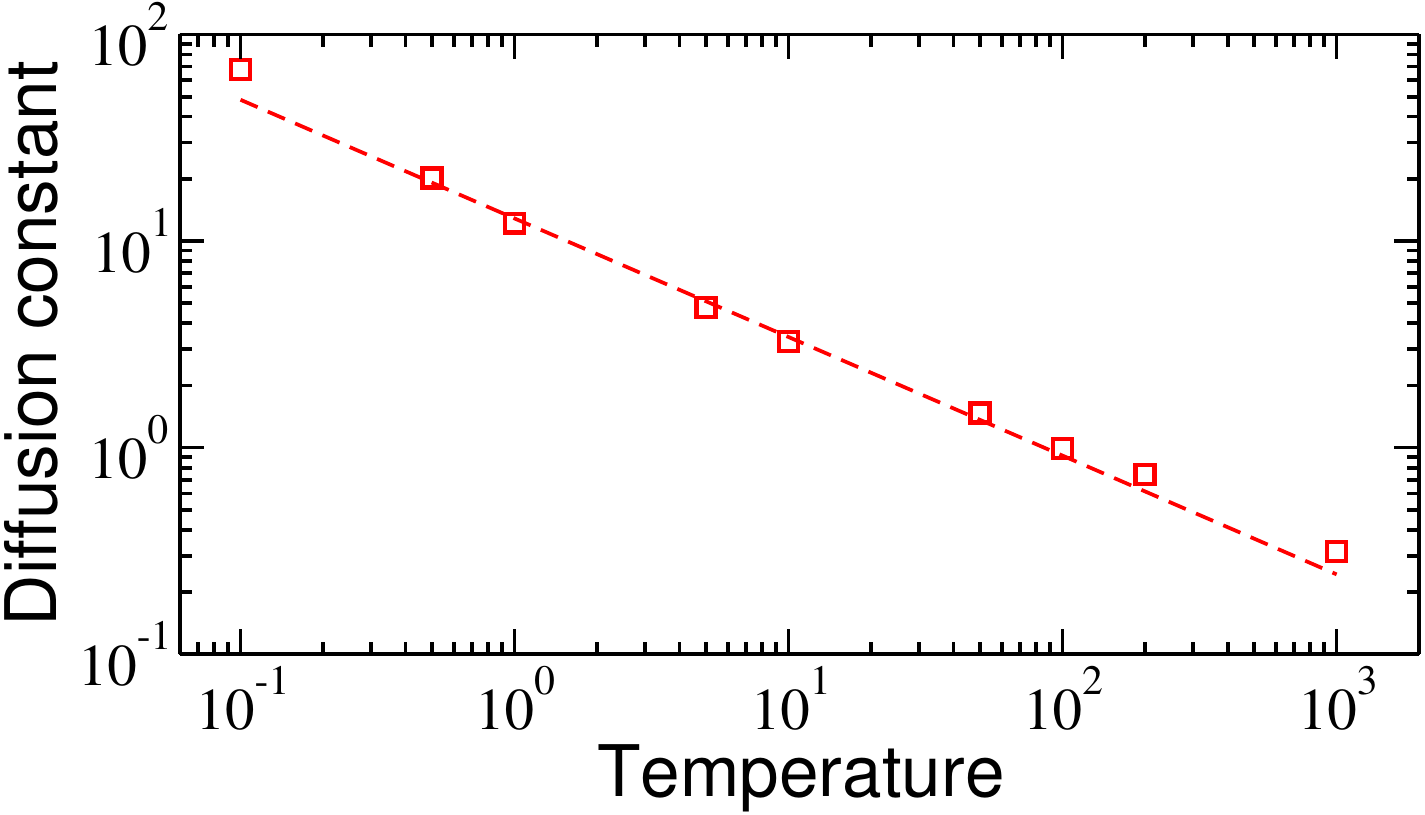}
\caption{(Color online) Temperature dependence of the spreading  
diffusion coefficient $D$ for an harmonic chain $(\beta=0)$ with pure diagonal
coupling ($\chi_E=1$ and $\chi_J=0$). The red line is a power law fit
with a function $D \propto T^{-\gamma}$ with $\gamma=0.58$. }
\label{f:fig5}
\end{center}
\end{figure}

\noindent From the above discussion it is clear that in the intermediate 
coupling regime the asymptotic  dynamics is diffusive. We now turn to  
analyzing in detail the properties of the diffusive spreading by 
a characterization of the diffusion constant $D$, defined as
\begin{equation}
  D = \lim_{t\rightarrow\infty} t^{-1/2}\Pi(t) 
\end{equation}
In Fig.~\ref{f:fig2} we compare the growth of $\Pi(t)$ for increasing 
temperatures $T$  and fixed values of the couplings $\chi_{E}=\chi_{J}=1$ in the 
intermediate regime. Interestingly, we find a  nonmonotonic behavior for the dependence 
of the diffusion constant on the temperature $T$.
More precisely, we observe a minimum located around $T=10$, indicating the 
presence of a slowed-down spreading dynamics at intermediate temperatures.  \\
%
%
%
\indent This phenomenon can be  illustrated more effectively by measuring the
average time $t_s$ it takes for the participation ratio to reach a certain threshold 
value $\Pi_s$. In this {\em temporal representation} a minimum in the diffusion 
constant $D$ corresponds to a maximum of the time $t_s$.
The value of $\Pi_s$ needs to be chosen in such a way as to avoid both the 
transient dynamics (typical of short times) and the saturation of $\Pi(t)$  due 
to the finite size  of the considered lattices (see again Fig.~\ref{f:fig2}). 
Accordingly, we have chosen  $\Pi_s=100$ for a chain of $L=1000$ sites. 
The dependence of $t_s$ on temperature is illustrated in Fig.~\ref{f:fig3} 
for increasing couplings.  For the lowest coupling considered, 
$t_{s}$ displays an initial growing stage, followed by a decrease at high temperatures
after a maximum, which becomes more and more distinguishable upon increasing 
the coupling (note the logarithmic scale on the $y$ axis in Fig.~\ref{f:fig3}).
Interestingly, this maximum appears to move towards smaller and  smaller temperatures
at increasing coupling strengths. This feature should be compared with the 
data displayed in the inset in Fig.~\ref{f:fig2}, illustrating a minimum of the 
transport coefficient associated with exciton transport. A stationary point at the 
same value of temperature $T \approx 10$ is indeed recovered in both mobility indicators,
$D$ and $t_{s}$.\\
\indent The sudden growth of $t_s$  in the low-temperature region 
for $\chi=2.5$ flags the presence of self-trapped, breather-like excitations, which
pin energy locally and thus slow down the relaxation process. 
This kind of localized states, however, are 
not present at the temperatures characterizing the maximum of $t_s$, since the 
strength of thermal fluctuations is too large to sustain coherent 
localized nonlinear vibrations~\cite{Davydov:1979,Cruzeiro-Hansson:1995hc}. \\
%
%
%
\indent The above analysis suggests that the nonmonotonicity displayed by transport coefficients 
with temperature is not related directly to nonlinear localization phenomena that 
pin energy down at high temperatures. It is therefore highly likely that the 
observed transport behavior is directly linked to the specific way the lattice vibrations
couple to the parameters entering the exciton Hamiltonian.  
In fact, in the limit case of spatially uncorrelated white noise, 
it has been long known that local and nonlocal perturbations of the
coherent Schr\"odinger dynamics can produce dramatically different behaviors 
for the spreading of an initially localized excitation~\cite{ScwharzerHaken1972}. 
In particular, it is known that purely local noise (in our picture, dynamical modulation of the 
site energies only) results in suppression of transport for large dephasing rates (a phenomenon 
often considered as an instance of the quantum Zeno 
effect~\cite{Misra1977zeno,rebentrost2009environment}).\\
\indent With these ideas in mind, we turned to examine the role of the effective noise 
acting on the free exciton dynamics as a consequence of the lattice thermal fluctuations.
In the same spirit as the analysis performed in Ref.~\cite{ScwharzerHaken1972}, 
we simplified the coupling between the exciton and the lattice by  studying separately diagonal 
(involving site energies) and off-diagonal (involving hopping rates) interactions.\\
\indent In Fig.~\ref{f:fig4} we compare the spreading diffusion constant 
of a system exhibiting only diagonal coupling ($\chi_J=0$) with the one 
corresponding to a pure off-diagonal coupling ($\chi_E=0$).
Interestingly, the nonmonotonic behavior of $D$ is present only in the latter case, 
while in the former we observe a monotonic decrease of $D$ with temperature.
This is precisely what happens in a master equation description {\em \`a la} Haken and 
Strobl~\cite{haken1973exactly} with nearest-neighbor Coulomb coupling, where 
\begin{equation}
\label{beccatiquesta}
D = 2\gamma_{1}a^{2} + \frac{a^{2}J^{2}}{\hbar^{2}(\gamma_{0} + 3\gamma_{1})}
\end{equation}
Here $\gamma_{0}$ and $\gamma_{1}$ are the diagonal (pure dephasing) and 
off-diagonal noise strengths, $J$ is the Coulomb hopping integral and $a$ is 
the lattice spacing.  The results of our simulations performed with $\chi_{J}=0$
are in agreement with the  prediction~\eqref{beccatiquesta} with $\gamma_{1}=0$,
that is, a value of $D$ which decreases monotonically with temperature ($\gamma_{0}$ in 
the language of Refs.~~\cite{haken1973exactly} and~\cite{rebentrost2009environment}.
In fact, the effect of  purely local perturbations of the 
excitonic energy is to produce a diffusion of the Schr\"odinger phases that 
inhibits quantum transport. Eventually, in the limit of infinite 
interaction (i.e. infinite temperature), the quantum
system remains frozen in the initial condition ($D=0$) as a consequence of the 
complete randomization of the phases.\\
%
%
%
\indent The situation changes when we let the hopping rates be modulated 
by the lattice dynamics ($\chi_{J}\neq 0$, {\em i.e.} $\gamma_{1} \neq 0$). 
As can be appreciated from Fig.~\ref{f:fig3},
the diffusion constant displays a minimum, in agreement with the general 
prediction~\eqref{beccatiquesta} and the transport becomes faster at 
increasing temperatures. In fact, since both phase and amplitude 
perturbations are now allowed for, the limit of infinite temperature 
corresponds to  an arbitrarily large diffusion constant.
This result can be interpreted as a recovery of classical 
amplitude diffusion in the infinite temperature limit.  
On more formal grounds, it can be shown that such classical reduction 
allows one to reduce a generalized Lindblad equation~\cite{ScwharzerHaken1972} 
for the exciton density matrix to a classical Fokker-Planck 
equation for the amplitude probability distribution. 
Although the non-Markovian nature of the effective noise and the explicit 
nonlinearity make the analytical calculation of 
$D(T,\chi_E,\chi_J)$ extremely difficult, the numerical results 
reported in Fig.~\ref{f:fig4} clearly indicate that classical diffusion is recovered 
also when one considers the full model with both local and nonlocal 
couplings.\\
\indent Interestingly, we remark that at low temperatures
pure-diagonal noise allows for the fastest transport. However, as non-diagonal 
noise results in a minimum, the situation reverses beyond a characteristic temperature
when $\chi_{J}$ is {\em switched on}. In this case, the model with fluctuating coupling 
strengths becomes the one affording  more rapid spreading at high temperatures. \\
\indent It is interesting to note that extended vibrational modes are required 
in order to observe a nonmonotonic behavior of the diffusion constant, signaling 
a nontrivial coupling between exciton spreading and {\em collective} modes of the 
underlying lattice. This can be appreciated by comparing our analysis with  
the results presented in Ref.~\cite{Troisi:2006fk}, where the hopping rates in the tight-binding 
exciton Hamiltonian are modulated by the dynamics of a set of {\em uncoupled}
harmonic oscillators. In this case, the authors report values of the diffusion 
coefficient that decrease monotonically with temperature. This is possibly a  consequence
of the absence of coupling between the oscillators providing the noise. 
Alternatively, they might be just exploring the low-temperature regime, as defined 
by their parameters.\\ 
\indent From Fig.~\ref{f:fig4} one can also argue  that 
the effective interaction experienced by the exciton for finite temperatures
can not be mapped onto a spatially and temporally
uncorrelated noise as in~\cite{ScwharzerHaken1972}. Specifically, we find that the
spreading problem for pure local exciton-lattice interactions is ruled by a
nontrivial power-law decay $D(T)\sim T^{-\gamma}$ with $\gamma\approx 0.3$
(data in Fig.~\ref{f:fig4}), whereas $\gamma=1$ for white noise~\cite{ScwharzerHaken1972}.
In the absence of nonlinearity in the lattice Hamiltonian ($\beta=0$), the
characteristic exponent is found to be close to $\gamma  \simeq 0.6$, as
shown in Fig.~\ref{f:fig5}. We therefore identify two different sources of slowing down
in the transport, that are characteristic of the explicit lattice dynamics. 
The first one is due  to the presence of spatio-temporal correlations in the lattice system. 
The second one is associated with explicit nonlinear terms in the lattice pairwise 
potential energies $V\left(u_n,u_m\right)$.\\
{\bf here put some additional consideration on the effect of the $\beta$ nonlinearity.}
%
%
\indent Overall, the above detailed analysis allows us to conclude that the 
high-temperature limit of excitonic systems interacting with noisy environments  
is crucially determined by the specific properties of the exciton-lattice coupling and 
by the nature of the spatio-temporal correlations that characterize the noise-providing 
underlying lattice. 
In particular, the presence of non-diagonal coupling is sufficient to suppress localization  
at high temperatures, which can be regarded as the semiclassical counterpart of 
the quantum Zeno effect observed in quantum master equation approaches~\cite{Rebentrost2009environment}.\\
\indent An interesting consequence of the above reasoning is that the standard scenario for 
noise-assisted quantum transfer efficiency~\cite{Chin2010,rebentrost2009environment} 
may display  novel  structures when passing from local pure dephasing noise to more 
realistic models of coupling including amplitude-affecting terms in the non-Hermitian 
part of the Hamiltonian. We thus turn now to discussing the implications of our explicit-noise approach
for the efficiency problem.
%
%
%
%
\section{Exciton energy transport efficiency}\label{sec:efficiency}

A quantum excitation such an exciton has an intrinsic lifetime, which is 
dictated by the recombination rate $\gamma_{r}$ associated with the specificities 
of the environment. For example, in light-harvesting systems  $\gamma_{r}$ is 
estimated to be about 1 exciton per nanosecond~\cite{rebentrost2009environment}.
The quantum excitation is therefore damped as it spreads through the system 
following its excitation. 
It is interesting to  provide a  measure of {\em efficiency} associated  with 
the transport of an exciton. This can be done by requiring that a sink 
exists at some specific location in the system ({\em e.g.} allowing the excitation to
be transferred to a neighboring equivalent system) and evaluating the probability that 
the exciton exits through the sink rather than decaying {\em non-specifically} due
to recombination mechanisms.\\
\indent In a master equation description, a recombination probability and a sink appear as 
non-Hermitian terms in the time-evolution operator for the exciton density matrix. 
Similarly, in our approach we ought to add damping terms to the equations of motion~\eqref{EOMexciton}.
More precisely, we consider a chain where the site $k$ is identified as a sink, characterized
by a trapping rate $\G \gg \gamma_{r}$. Therefore, the modified equations of motion read
\be
\label{EOMexcitonREC}
i\hbar\frac{{\rm d} b_n}{{\rm d} t} = - \epsilon_n b_n 
                                      - J_n (b_{n+1} + b_{n-1}) 
                                      - i \hbar(\gamma_{r} + \delta_{nk}\Gamma) b_{n}
\ee
Along the lines of previous studies, such as Refs.~\cite{rebentrost2009environment},~\cite{Novo:2013kx}
and~\cite{Pelzer:2013ud}, we use eqs.~\eqref{EOMexcitonREC} to investigate 
the {\em competition} between the two mechanisms of exciton destruction, namely
generic recombination ($\gamma_{r}$) and exit through a specific {\em channel} ($\Gamma$). 
The whole idea is that an {\em efficient} transport is maximally effective in channeling 
the exciton rapidly through the specified exit site against the generic degradation
due to recombination. For example, this might reflect an exciton leaving a light-harvesting
complex through a specific pigment connected to the reaction center.
Moreover, as done in the above-cited studies, we 
take the unperturbed exciton site energies $E^0_n$ in Eq.~\eref{EOMexcitonREC} as random,
which has the well-known effect of inducing spatial localization of the exciton at zero 
temperature. \\
%
\begin{figure}[t!]   
\begin{center}
\includegraphics[width=0.7\textwidth,clip]{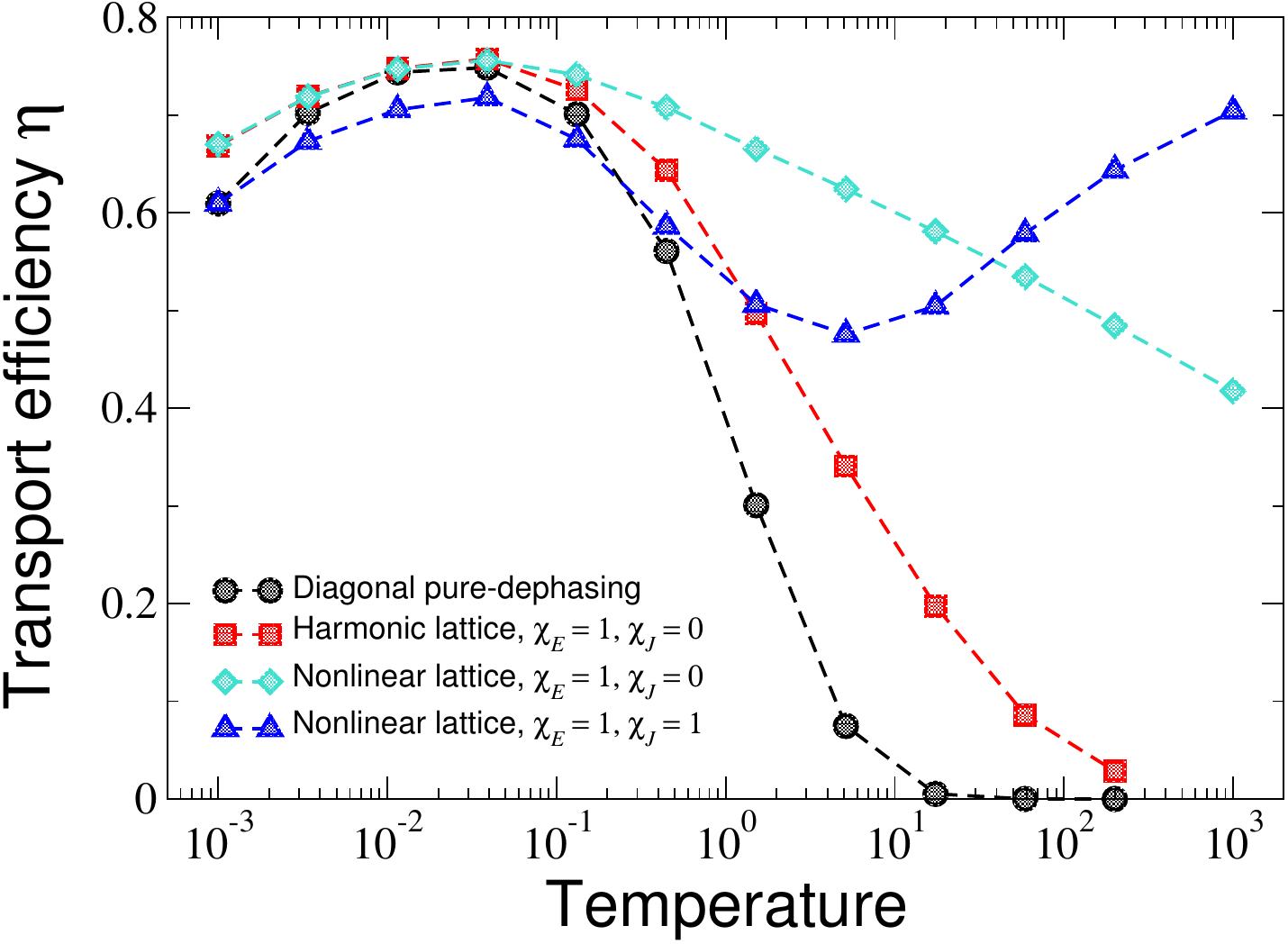}
\caption{(Color online) Exciton transport efficiency versus temperature 
in different kinds of lattices of length $L=200$. 
The recombination constant and the sink dissipation are,
respectively, $\gamma_r=10^{-4}$ and $\Gamma=10^{-1}$. 
The local unperturbed exciton energies 
$E_n^0$ are randomly chosen from  a uniform distribution in the interval
$[-0.5\, ;\,0.5 ]$. The efficiency $\eta$ is obtained by averaging over 20
independent realizations of the static disorder on the site energies 
$E_n^0$ and of the dynamical lattice noise.}
\label{f:figeff}
 \end{center}
\end{figure}
%
\indent The incorporation of non-Hermitian terms implies a loss of norm as time passes, 
until eventually the norm of the exciton wave-function reaches zero for infinite
times. Accordingly, a measure of transport efficiency can be computed in the following fashion
\begin{equation}\label{eq:eff}
\eta = 2 \int_0^\infty |b_{k}(t)|^{2}\,dt
\end{equation}
One can easily prove that $\eta$ is finite and takes values in
the interval $[0\,;\,1]$ as a consequence of the conservation of the total
excitonic amplitude when $\gamma_{r} = \Gamma = 0$. 
The integral in Eq.~\ref{eq:eff} can be computed numerically
to any desired accuracy,  as controlled by the amount of
total exciton norm $N(t) = \sum_{i}|b_{i}(t)|^{2}$ left in the system at time $t$.
In our simulations we integrated equations~\eqref{EOMexcitonREC}  
until the survival probability reached the threshold value of $10^{-5}$.\\
\indent In Fig.~\ref{f:figeff} we compare the transfer efficiency of our chain
model with the one corresponding to a quantum dynamics in the presence of
pure-dephasing (PD) noise~\cite{ScwharzerHaken1972,rebentrost2009environment}. 
The lattice is initially thermalized at temperature $T$ with the same procedure 
described in the introduction of Section~\ref{results}, 
while the PD dynamics 
describes the interaction of the exciton with a classical incoherent external
field (no explicit lattice in this case) that induces decoherence on the quantum system.
This is accomplished by specializing Eqs.~(\ref{renormenergy}) and (\ref{renormJ}) to
\be
\epsilon_n=E_n^0+\sqrt{2T}\xi_n(t), \qquad J_n=J_n^0
\ee
where $\xi_n(t)$ is a Gaussian white noise with zero mean and unit 
variance and $T$ effectively accounts for the dephasing rate of the process.  
In both cases the exciton is initially injected at one side of the chain, 
while a trap is located  at the opposite end. 
Interestingly,  in the low-temperature region the two systems display
qualitatively  the same behavior, namely a disruption of disorder-induced localization 
due to increasing dephasing (PD model), {\em i.e.} increasing 
thermal fluctuations of the underlying lattice in our scheme. 
However, at higher temperatures the efficiency of transport in the presence
of explicit ({\em i.e.} produced by the lattice) 
noise turns out to depend on the details of the underlying 
lattice dynamics. The first remarkable result is that nonlinear correlations 
in the lattice dynamics boost the efficiency in the Zeno-effect region (see diamonds
versus squares in Fig.~\ref{f:figeff}). This finding, although for somewhat 
different physical reasons, is in line with recent results where the importance of 
spatial correlations was demonstrated for transport in a PD-like model~\cite{Pelzer:2013ud}.
Furthermore, we note that such a behavior is consistent with 
the different scaling exponents  of the diffusion constant $D$ 
discussed in Section~\ref{results}.\\
\indent The second important finding is that non-diagonal noise ({\em i.e.} non-zero coupling 
between the hopping rates in the exciton Hamiltonian and the lattice) suppresses
the Zeno drop in the efficiency at high temperatures. This is in good agreement with 
the prediction of eq.~\eqref{beccatiquesta}, namely that the diffusion coefficient should 
be non-monotonic with temperature ({\em i.e.} non-diagonal noise, $\gamma_{1}$). 
We recall that in this regime, 
the master equation for the density matrix turns into a classical diffusion equation,
which should guarantee diffusive transport~\cite{schwarzer1972moments}, albeit possibly 
with a diffusion coefficient that decreases with temperature~\cite{Troisi:2006fk}.

%
\section{Conclusions}\label{sec:concl}
%
%
%
%
%
%
%
%
%
%

In this paper we have studied a model describing the dynamics of a quantum
excitation that propagates in a system at finite temperature. In our scheme, 
the quantum evolution is dictated by a tight-binding (TB) Hamiltonian, whose 
matrix elements are functions of the classical coordinates of an underlying 
one-dimensional lattice. We refer to this setting as an open quantum system with 
an {\em explicit} environment, which provides a direct source of noise to the 
quantum excitation, endowed with specific spatio-temporal correlations. 
While the main ideas behind this modeling scheme are not new (see for example 
Ref.~\cite{Davydov:1962vn}), our study contains important elements of novelty. 
Notably, we explicitly focussed on spotlighting the signatures
of non-trivial spatio-temporal correlations of nonlinear origin expressed 
by the underlying lattice. Furthermore, we conducted an original investigation 
of the relative role of diagonal and non-diagonal exciton-phonon couplings.
In both cases, we uncovered a rich phenomenology, prompting new directions of 
investigation.\\
\indent We first examined the spreading of an initially localized excitation. 
Our results show that the very first stage of the propagation is faster than ballistic, 
up to a time of the order of $\hbar/(2J)$, $J$ being the magnitude of 
the hopping integrals in the TB Hamiltonian. The subsequent time evolution is 
characterized by a transition to a ballistic stage followed by a crossover 
to an asymptotic diffusive regime, which appears at earlier and earlier times 
as the exciton-phonon coupling strength is increased. However, for large values of 
the coupling the picture changes dramatically,
as a self-trapped state of nonlinear character sets in after the first super-ballistic
spreading. The result is that the transport is completely suppressed in this regime 
as a sheer nonlinear effect.\\
\indent An analysis of the diffusion coefficient $D$ at intermediate exciton-phonon 
couplings unveils a striking non-monotonic behavior of $D$ as a function of temperature, 
provided the lattice is actively modulating the hopping integrals in the TB Hamiltonian. 
This effect, reported here for the first time in the presence of an explicit 
environment, agrees with a long-known prediction made on the basis of a generalized
master equation for the one-particle density matrix 
containing both dephasing and amplitude-affecting operators~\cite{ScwharzerHaken1972}.
Intriguingly, we find that  diffusive transport is faster at low temperatures with 
pure-diagonal noise ({\em i.e.} only on the site energies), 
but adding non-diagonal noise makes spreading faster at high 
temperatures.\\
\indent Importantly, our results on the diffusive regime at intermediate coupling 
flag a nontrivial interconnection between exciton spreading and collective (hydrodynamic) 
vibrational modes of the underlying lattice. More precisely, our findings strongly 
suggest that the observed non-monotonic behavior of the diffusion constant versus temperature 
is related to the presence of long-wavelength acoustic modes. This conclusion is 
reinforced by a comparison with the results of Troisi and Orlandi obtained
in a similar semi-classical model with purely off-diagonal dynamical disorder~\cite{Troisi:2006fk}.  
In fact, they found a monotonic decrease of $D$ with temperature in a model that lacks
collective vibrational modes by construction, as
their TB Hamiltonian is modulated by the dynamics of an ensemble of {\em independent}, disconnected
classical oscillators. In fact, the signatures of long-wavelength hydrodynamic modes 
are clearly recognizable in the equilibrium power spectra $S(k,\omega)$ 
of the exciton-coupled lattice, as shown in Figs.~\ref{f:fig0a}.
This strongly suggests that coupling to an {\em extended} dynamical system, such as the one 
we consider here, might be a necessary condition to obtain non-monotonic transport with temperature, 
in agreement with~\cite{ScwharzerHaken1972}. 
\\
\indent Our results on the role of the lattice in the spreading properties of a quantum 
excitation show that the presence of non-diagonal coupling is sufficient to suppress 
Zeno-like localization at high temperature. 
To shed further light into this phenomenon, 
we then computed a measure of quantum efficiency for different choices of the 
chain parameters. Our results clearly confirm that, when the hopping rates in the 
TB Hamiltonian  are explicitly modulated by the lattice dynamics, the transport 
efficiency is no longer quenched at high temperature, as observed by 
some authors in the absence of an 
{\em explicit} environment~\cite{Chin2010,rebentrost2009environment}.
Moreover, we find that nonlinearity in the lattice dynamics exerts a powerful
boosting action on the efficiency at high temperatures, confirming recent results 
on the importance of spatial and dynamical correlation patterns 
within the noise bath~\cite{Pelzer:2013ud}.\\
\indent Overall, the results presented in this paper allow us to conclude that the 
properties of excitonic systems interacting with noisy environments  
are subtly shaped by the specific properties of the exciton-phonon coupling and by
the nature of the spatio-temporal dynamical 
correlations that characterize the underlying lattice.
It would be extremely interesting to extend the formalism presented here 
to more complex systems, such as light-harvesting complexes, which are 
widely studied in the context of quantum biology~\cite{:2014fk,Huelga:2013uq}.

\section*{Acknowledgements}
The authors acknowledge financial support from the  EU FP7 project PAPETS (GA 323901).
YO and OB thank the support from Funda\c{c}\~{a}o para a Ci\^{e}ncia e a Tecnologia (Portugal), 
namely through programmes PTDC/POPH and projects PEst-OE/EGE/UI0491/2013, 
PEst-OE/EEI/LA0008/2013, UID/EEA/50008/2013, IT/QuSim and CRUP-CPU/CQVibes, 
partially funded by EU FEDER and from the EU FP7 project LANDAUER (GA 318287).

\section*{References}


\begin{thebibliography}{10}
\expandafter\ifx\csname url\endcsname\relax
  \def\url#1{{\tt #1}}\fi
\expandafter\ifx\csname urlprefix\endcsname\relax\def\urlprefix{URL }\fi
\providecommand{\eprint}[2][]{\url{#2}}

\bibitem{Law2004}
Law M, Goldberger J and Yang P 2004 {\em Annual Review of Materials Research\/}
  {\bf 34} 83--122

\bibitem{Yanson:1998ly}
Yanson A~I, Bollinger G~R, van~den Brom H~E, Agrait N and van Ruitenbeek J~M
  1998 {\em Nature\/} {\bf 395} 783--785

\bibitem{Roati2008}
Roati G, D'Errico C, Fallani L, Fattori M, Fort C, Zaccanti M, Modugno G,
  Modugno M and Inguscio M 2008 {\em Nature\/} {\bf 453} 895--898

\bibitem{Perebeinos2004}
Perebeinos V, Tersoff J and Avouris P 2004 {\em Physical Review Letters\/} {\bf
  92} 257402--1

\bibitem{Wang:2007bh}
Wang F, Cho D~J, Kessler B, Deslippe J, Schuck P~J, Louie S~G, Zettl A, Heinz
  T~F and Shen Y~R 2007 {\em Phys. Rev. Lett.\/} {\bf 99}(22) 227401

\bibitem{Zhao:2006qf}
Zhao H, Mazumdar S, Sheng C~X, Tong M and Vardeny Z~V 2006 {\em Phys. Rev. B\/}
  {\bf 73}(7) 075403

\bibitem{Lee:2007zr}
Lee J, Hernandez P, Lee J, Govorov A~O and Kotov N~A 2007 {\em Nature
  materials\/} {\bf 6} 291--295

\bibitem{Scholes2006}
Scholes G~D and Rumbles G 2006 {\em Nature materials\/} {\bf 5} 683--696

\bibitem{Blankenship:2002ve}
Blankenship R~E 2002 {\em Molecular mechanisms of photosynthesis\/} (Blackwell
  Science)

\bibitem{Amerongen:2000ly}
van Amerongen H, van Grondelle R and Valkunas L 2000 {\em Photosynthetic
  Excitons\/} (World Scientific)

\bibitem{Frenkel:1931zr}
Frenkel J 1931 {\em Phys. Rev.\/} {\bf 37}(1) 17--44

\bibitem{Davydov:1948ys}
Davydov A~S 1948 {\em Zhur. Eksptl. i Teoret. Fiz.\/} {\bf 18} 210--218

\bibitem{Engel2007}
Engel G~S, Calhoun T~R, Read E~L, Ahn T~K, Mancal T, Cheng Y~C, Blankenship R~E
  and Fleming G~R 2007 {\em Nature\/} {\bf 446} 782--6

\bibitem{Read2009}
Read E~L, Lee H and Fleming G~R 2009 {\em Photosynthesis Research\/} {\bf 101}
  233--243

\bibitem{Collini2010}
Collini E, Wong C~Y, Wilk K~E, Curmi P~M~G, Brumer P and Scholes G~D 2010 {\em
  Nature\/} {\bf 463} 644--647

\bibitem{Romero:2014ve}
Romero E, Augulis R, Novoderezhkin V~I, Ferretti M, Thieme J, Zigmantas D and
  van Grondelle R 2014 {\em Nat Phys\/} {\bf 10} 676--682

\bibitem{Collini2009}
Collini E and Scholes G~D 2009 {\em Science (New York, N.Y.)\/} {\bf 323}
  369--373

\bibitem{Chin2013}
Chin a~W, Prior J, Rosenbach R, Caycedo-Soler F, Huelga S~F and Plenio M~B 2013
  {\em Nature Physics\/} {\bf 9} 113--118

\bibitem{Chin2010}
Chin a~W, Datta a, Caruso F, Huelga S~F and Plenio M~B 2010 {\em New Journal of
  Physics\/} {\bf 12} 065002

\bibitem{rebentrost2009environment}
Rebentrost P, Mohseni M, Kassal I, Lloyd S and Aspuru-Guzik A 2009 {\em New
  Journal of Physics\/} {\bf 11} 033003

\bibitem{OReilly2014}
O'Reilly E~J and Olaya-Castro A 2014 {\em Nature communications\/} {\bf 5} 3012

\bibitem{Tiwari2013}
Tiwari V, Peters W~K and Jonas D~M 2013 {\em Proceedings of the National
  Academy of Sciences of the United States of America\/} {\bf 110} 1203--8

\bibitem{Mennucci2011}
Mennucci B and Curutchet C 2011 {\em Physical chemistry chemical physics :
  PCCP\/} {\bf 13} 11538--11550

\bibitem{Plenio:2008fk}
Plenio M~B and Huelga S~F 2008 {\em New Journal of Physics\/} {\bf 10} 113019

\bibitem{Skourtis:2007kx}
Skourtis S~S and Beratan D~N 2007 {\em Science\/} {\bf 316} 703--704

\bibitem{Lee2007}
Lee H, Cheng Y~C and Fleming G~R 2007 {\em Science (New York, N.Y.)\/} {\bf
  316} 1462--5

\bibitem{Wang:2007vn}
Wang H, Lin S, Allen J~P, Williams J~C, Blankert S, Laser C and Woodbury N~W
  2007 {\em Science\/} {\bf 316} 747--750

\bibitem{Adolphs2006}
Adolphs J and Renger T 2006 {\em Biophysical journal\/} {\bf 91} 2778--97

\bibitem{Saito:2007qf}
Saito K and Dhar A 2007 {\em Phys. Rev. Lett.\/} {\bf 99}(18) 180601

\bibitem{Chetverikov:2009ys}
Chetverikov A~P, Ebeling W and Velarde M~G 2009 {\em The European Physical
  Journal B\/} {\bf 70} 217--227

\bibitem{Komarnicki:2003uq}
Komarnicki S and Hennig D 2003 {\em Journal of Physics: Condensed Matter\/}
  {\bf 15} 441

\bibitem{Leitner:2001ys}
Leitner D~M 2001 {\em Phys. Rev. B\/} {\bf 64} 094201

\bibitem{Platero2004}
Platero G and Aguado R 2004 {\em Physics Reports\/} {\bf 395} 1--157

\bibitem{Mulansky:2013zr}
Mulansky M and Pikovsky A 2013 {\em New Journal of Physics\/} {\bf 15} 053015

\bibitem{Ivanchenko:2011ly}
Ivanchenko M~V, Laptyeva T~V and Flach S 2011 {\em Phys. Rev. Lett.\/} {\bf
  107}(24) 240602

\bibitem{Flach2010}
Flach S 2010 {\em Chemical Physics\/} {\bf 375} 548--556

\bibitem{Garcia-Mata2009}
Garc\'{\i}a-Mata I and Shepelyansky D~L 2009 {\em Physical Review E -
  Statistical, Nonlinear, and Soft Matter Physics\/} {\bf 79} 026205

\bibitem{Kopidakis2008}
Kopidakis G, Komineas S, Flach S and Aubry S 2008 {\em Physical Review
  Letters\/} {\bf 100} 084103

\bibitem{Vlaming:2012qy}
Vlaming S~M and Silbey R~J 2012 {\em The Journal of chemical physics\/} {\bf
  136} 055102

\bibitem{Chen:2010uq}
Chen X and Silbey R~J 2010 {\em The Journal of chemical physics\/} {\bf 132}
  204503

\bibitem{Arago:2015bh}
Arag\'o J and Troisi A 2015 {\em Phys. Rev. Lett.\/} {\bf 114}(2) 026402

\bibitem{Troisi2011}
Troisi A 2011 {\em Chemical Society reviews\/} {\bf 40} 2347--2358

\bibitem{ScwharzerHaken1972}
Schwarzer E and Haken H 1972 {\em Physics Letters A\/} {\bf 42} 317--318

\bibitem{haken1973exactly}
Haken H and Strobl G 1973 {\em Zeitschrift f{\"u}r Physik\/} {\bf 262} 135--148

\bibitem{Troisi:2006fk}
Troisi A and Orlandi G 2006 {\em Phys. Rev. Lett.\/} {\bf 96}(8) 086601

\bibitem{Breuer:2002ve}
Breuer H~P and Petruccione F 2002 {\em The Theory of Open Quantum Systems\/}
  (Oxford University Press)

\bibitem{Rivas:2011qf}
Rivas A and Huelga S~F 2011 {\em Open Quantum Systems: An Introduction\/}
  (Springer)

\bibitem{schwarzer1972moments}
Schwarzer E and Haken H 1972 {\em Physics Letters A\/} {\bf 42} 317--318

\bibitem{Lindblad:1976tg}
Lindblad G 1976 {\em Commun. Math. Phys.\/} {\bf 48} 119

\bibitem{V.-Gorini:1976hc}
Gorini V, Kossakowski A and Sudarshan E~C~G 1976 {\em Journ. of Math. Phys.\/}
  {\bf 17} 821

\bibitem{Nakajima:1958bh}
Nakajima S 1958 {\em Progress of Theoretical Physics\/} {\bf 20} 948--959

\bibitem{Ishizaki:2009dq}
Ishizaki A and Fleming G~R 2009 {\em Proceedings of the National Academy of
  Sciences\/} {\bf 106} 17255--17260

\bibitem{Chen2011}
Chen X and Silbey R~J 2011 {\em Journal of Physical Chemistry B\/} {\bf 115}
  5499--5509

\bibitem{Prior:2010uq}
Prior J, Chin A~W, Huelga S~F and Plenio M~B 2010 {\em Phys. Rev. Lett.\/} {\bf
  105}(5) 050404

\bibitem{Boninsegna:2012kx}
Boninsegna L and Faccioli P 2012 {\em The Journal of Chemical Physics\/} {\bf
  136} 214111

\bibitem{Huo:2010uq}
Huo P and Coker D~F 2010 {\em The Journal of Chemical Physics\/} {\bf 133} --

\bibitem{Makri:1995cr}
Makri N and Makarov D~E 1995 {\em The Journal of Chemical Physics\/} {\bf 102}
  4600--4610

\bibitem{Kurnosov:2015vn}
Kurnosov A~A, Rubtsov I~V and Burin A~L 2015 {\em The Journal of Chemical
  Physics\/} {\bf 142} 011101

\bibitem{Shim2012}
Shim S, Rebentrost P, Valleau S and Aspuru-Guzik A 2012 {\em Biophysical
  journal\/} {\bf 102} 649--60

\bibitem{Gutierrez:2009nx}
Guti\'errez R, Caetano R~A, Woiczikowski B~P, Kubar T, Elstner M and Cuniberti
  G 2009 {\em Phys. Rev. Lett.\/} {\bf 102}(20) 208102

\bibitem{Cheung:2009oq}
Cheung D~L, McMahon D~P and Troisi A 2009 {\em The Journal of Physical
  Chemistry B\/} {\bf 113} 9393--9401

\bibitem{Viani:2014kl}
Viani L, Corbella M, Curutchet C, O'Reilly E~J, Olaya-Castro A and Mennucci B
  2014 {\em Physical Chemistry Chemical Physics\/} {\bf 16} 16302--16311

\bibitem{Hennig2001}
Hennig D 2001 {\em European Physical journal B\/} {\bf 381} 377--381

\bibitem{Mingaleev1999}
Mingaleev S~F, Christiansen P~L, Gaididei Y~B, Johansson M and Rasmussen K~O
  1999 {\em Journal of biological physics\/} {\bf 25} 41--63

\bibitem{Bittner2010}
Bittner E~R, Goj A~M and Burghardt I 2010 {\em Chemical Physics\/} {\bf 370}
  137--142

\bibitem{Freedman:2010kx}
Freedman H, Martel P and Cruzeiro L 2010 {\em Phys. Rev. B\/} {\bf 82}(17)
  174308

\bibitem{Davydov:1962vn}
Davydov A~S 1962 {\em Theory of molecular excitons\/} (McGraw-Hill)

\bibitem{Scott1992}
Scott A 1992 {\em Physics Reports\/} {\bf 217} 1--67

\bibitem{Holstein1959}
Holstein T 1959 {\em Annals of Physics\/} {\bf 8} 325--342

\bibitem{Kalosakas:1998bs}
Kalosakas G, Aubry S and Tsironis G~P 1998 {\em Phys. Rev. B\/} {\bf 58}
  3094--3104

\bibitem{Su:1979ij}
Su W~P, Schrieffer J~R and Heeger A~J 1979 {\em Phys. Rev. Lett.\/} {\bf
  42}(25) 1698--1701

\bibitem{E:1955hs}
E F, J P and S U 1955 {\em Los Alamos Rep. LA-1940\/}

\bibitem{Butt:2007aa}
Butt I~A and Wattis J~A 2007 {\em Physica D: Nonlinear Phenomena\/} {\bf 231}
  165--179

\bibitem{LLP03}
Lepri S, Livi R and Politi A 2003 {\em Phys. Rep.\/} {\bf 377} 1

\bibitem{Fleischer:2003fv}
Fleischer J~W, Segev M, Efremidis N~K and Christodoulides D~N 2003 {\em
  Nature\/} {\bf 422} 147--150

\bibitem{Franzosi:2011dz}
Franzosi R, Livi R, Oppo G~L and Politi A 2011 {\em Nonlinearity\/} {\bf 24}
  R89

\bibitem{Flach2008}
Flach S and Gorbach A~V 2008 {\em Physics Reports\/} {\bf 467} 1--116

\bibitem{Trombettoni2001}
Trombettoni A and Smerzi A 2001 {\em Physical Review Letters\/} {\bf 86}
  2353--2356

\bibitem{Davydov:1979}
Davydov A~S 1979 {\em Physica scripta\/} {\bf 20} 387

\bibitem{Cruzeiro-Hansson:1995hc}
Cruzeiro-Hansson L and Kenkre V 1995 {\em Physics Letters A\/} {\bf 203}
  362--366

\bibitem{Misra1977zeno}
Misra B and Sudarshan E~C~G 1977 {\em Journal of Mathematical Physics\/} {\bf
  18} 756--763

\bibitem{Novo:2013kx}
Novo L, Mohseni M and Omar Y 2013 {\em arXiv preprint arXiv:1312.6989\/}

\bibitem{Pelzer:2013ud}
Pelzer K~M, Fidler A~F, Griffin G~B, Gray S~K and Engel G~S 2013 {\em New
  Journal of Physics\/} {\bf 15} 095019

\bibitem{:2014fk}
Mohseni M, Omar Y, Engel G and Plenio M~B (eds) 2014 {\em Quantum Effects in
  Biology\/} (Cambridge University Press)

\bibitem{Huelga:2013uq}
Huelga S~F and Plenio M~B 2013 {\em Contemporary Physics\/} {\bf 54} 181--207

\end{thebibliography}

\providecommand{\newblock}{}

\end{document}